\documentclass[pdflatex,sn-basic]{sn-jnl}

\usepackage{graphicx}%
\usepackage{multirow}%
\usepackage{amsmath,amssymb,amsfonts}%
\usepackage{amsthm}%
\usepackage{mathrsfs}%
\usepackage[title]{appendix}%
\usepackage{xcolor}%
\usepackage{textcomp}%
\usepackage{manyfoot}%
\usepackage{booktabs}%
\usepackage{algorithm}%
\usepackage{algorithmicx}%
\usepackage{algpseudocode}%
\usepackage{listings}%

\theoremstyle{thmstyleone}%
\theoremstyle{thmstyletwo}%

\theoremstyle{thmstylethree}%
\begin{document}

\title[The Polar Stratosphere of Jupiter]{The Polar Stratosphere of Jupiter}


\author*[1]{\fnm{V.} \sur{Hue}}\email{vincent.hue@lam.fr}

\author[2,3]{\fnm{T.} \sur{Cavalié}}

\author[4]{\fnm{J. A.} \sur{Sinclair}}

\author[5]{\fnm{X.} \sur{Zhang}}

\author[6]{\fnm{B.} \sur{Benmahi}}

\author[3]{\fnm{P.} \sur{Rodríguez-Ovalle}}

\author[7]{\fnm{R. S.} \sur{Giles}}

\author[8]{\fnm{T. S.} \sur{Stallard}}

\author[9]{\fnm{R. E.} \sur{Johnson}}

\author[2]{\fnm{M.} \sur{Dobrijevic}}

\author[3]{\fnm{T.} \sur{Fouchet}}

\author[7]{\fnm{T. K.} \sur{Greathouse}}

\author[6]{\fnm{D. C.} \sur{Grodent}}

\author[10]{\fnm{R.} \sur{Hueso}}

\author[1]{\fnm{O.} \sur{Mousis}}

\author[11]{\fnm{C. A.} \sur{Nixon$^{\text{11}}$}}

\affil*[1]{Aix-Marseille Université, CNRS, CNES, Institut Origines, LAM, Marseille, France}

\affil[2]{Laboratoire d’Astrophysique de Bordeaux, Univ. Bordeaux, CNRS, B18N, allée Geoffroy Saint-Hilaire, 33615 Pessac, France}

\affil[3]{LESIA, Observatoire de Paris, Université PSL, Sorbonne Université, Université Paris Cité, CNRS, 5 place Jules Janssen, 92195 Meudon, France}

\affil[4]{Jet Propulsion Laboratory, California Institute of Technology, 4800 Oak Grove Dr, Pasadena, CA 91109, United States}

\affil[5]{Department of Earth and Planetary Sciences, University of California Santa Cruz, Santa Cruz, CA 95064, United States}

\affil[6]{Laboratory for Planetary and Atmospheric Physics, STAR Institute, University of Liege, Liege, Belgium}

\affil[7]{Space Science and Engineering Division, Southwest Research Institute, San Antonio, TX, USA}

\affil[8]{Department of Mathematics, Physics and Electrical Engineering, Northumbria University, Newcastle upon Tyne, UK}

\affil[9]{Department of Physics, Aberystwyth University, Ceredigion, UK}

\affil[10]{Escuela de Ingeniería de Bilbao, Universidad del País Vasco, UPV/EHU, Bilbao, Spain}

\affil[11]{Solar System Exploration Division, NASA Goddard Space Flight Center, Greenbelt, MD 20771, USA}


\abstract{Observations of the Jovian upper atmosphere at high latitudes in the UV, IR and mm/sub-mm all indicate that the chemical distributions and thermal structure are broadly influenced by auroral particle precipitations. Mid-IR and UV observations have shown that several light hydrocarbons (up to 6 carbon atoms) have altered abundances near Jupiter's main auroral ovals. Ion-neutral reactions influence the hydrocarbon chemistry, with light hydrocarbons produced in the upper stratosphere, and heavier hydrocarbons as well as aerosols produced in the lower stratosphere. One consequence of the magnetosphere-ionosphere coupling is the existence of ionospheric jets that propagate into the neutral middle stratosphere, likely acting as a dynamical barrier to the aurora-produced species. As the ionospheric jets and the background atmosphere do not co-rotate at the same rate, this creates a complex system where chemistry and dynamics are intertwined. The ion-neutral reactions produce species with a spatial distribution following the SIII longitude system in the upper stratosphere. As these species sediment down to the lower stratosphere, and because of the progressive dynamical decoupling between the ionospheric flows and the background atmosphere, the spatial distribution of the auroral-related species progressively follows a zonal distribution with increasing pressures that ultimately produces a system of polar and subpolar hazes that extends down to the bottom of the stratosphere. This paper reviews the most recent work addressing different aspects of this environment.}

\maketitle

\section{Introduction}
\label{s:intro}

The four giant planets each possess a hydrogen/helium-dominated atmosphere with a variable amount of heavy elements. Remote-sensing observations allow us to derive a wealth of information on their stratospheres, from the chemical distributions to their thermal structure and dynamics. The stratosphere lies between the deep convective atmosphere, delimited by the temperature minimum around 100\,mbar \citep{Robinson2014}, and the external environment beyond the sub-$\mu$bar pressure levels. The chemical distributions are observable clues that reflect the internal chemical processes, but also carry the signature of the interaction with the external environment (through, e.g., external supply, auroral precipitations, etc.), and are influenced by the atmospheric dynamics. 

Understanding these systems as a whole is complex and requires analyzing each piece of the puzzle individually. Monitoring the chemical distributions is one such piece. Oxygen species broadly derive from external sources (comets, interplanetary dust particles, and interaction with moons/rings) with a smaller contribution from the lower atmosphere. Hydrocarbons however derive from methane photochemistry, and their distribution is controlled by diffusive equilibrium, and seasonal forcing. Observations, when interpreted alongside atmospheric models, provide information on the atmospheric chemical, thermal and dynamical properties.


Stellar and solar occultations in the far-UV allow the thermal profiles and composition of the upper neutral atmosphere to be constrained, including density profiles of molecular hydrogen, methane, ethane and acetylene \citep{Broadfoot1981, Festou1981, Greathouse2010}. Radio occultations provide another way to sound the giant planet upper atmospheres. As the signal is progressively refracted by the atmosphere, the change in received radio frequency can be used to infer ionospheric properties (electron density) as well as neutral atmospheric properties such as density, pressure, and temperature \citep{Lindal1981, Kliore2004}. Mid-infrared (IR) (5-25 $\mu$m) emission spectra of the gas giants are characterized by a high density of rovibrational lines from various light gases (CH$_4$, NH$_3$, PH$_3$, C$_2$H$_2$, C$_2$H$_4$, C$_2$H$_6$), and higher-order hydrocarbons such as C$_3$H$_8$, CH$_3$C$_2$H, C$_4$H$_2$, C$_6$H$_6$ \citep[e.g.][]{Fletcher2009b, Guerlet2009, Guerlet2010, Sinclair2019, Roman2023}. At longer wavelengths, in the millimeter (mm)/sub-millimeter range, individual rotational lines can be resolved due to the lower spectral line density combined with the very high resolving power offered by heterodyne spectroscopy. The resolved lineshapes in the mm/sub-mm range and at high-spectral resolution in the mid-IR provide constraints regarding the vertical distribution of the observed species. 

Many of the aforementioned observations have been carried out via constantly evolving ground- and space-based observing capabilities. In the mid-IR, a large range of instruments and telescopes have been used to obtain spectroscopic and medium- to narrow-band filtered radiometric observations of Jupiter over the last four decades. Ground-based observatories have allowed the measurement of hydrocarbon emissions in the mid-IR at spectral resolutions $R$ on the order of 10$^{3}$ ($R$ = $\lambda$/${\Delta \lambda}$), \citep[e.g.][]{Gillett1969, Ridgway1974, Encrenaz1978, Sada1998, Moses2005a}. Higher spectral resolutions ($R$ $>$ 10$^{4}$) have been achieved using IR heterodyne spectroscopy techniques \citep{Kostiuk1989, Livengood1993}, or with Fourier Transform Spectrometer \citep{Ridgway1984} and grating spectrographs \citep{Lacy2002}. Space-based instruments, such as the Infrared Interferometer Spectrometer and Radiometer (IRIS) on Voyager \citep{Hanel1979a, Hanel1979b}, the Short-Wavelength Spectrometer (SWS) on the Infrared Space Observatory (ISO) \citep{DeGraauw1996}, or the Composite Infrared Spectrometer (CIRS) on Cassini-Huygens \citep{Kunde1996, Kunde2004} provided mid-IR spectroscopic observations of Jupiter at high sensitivity (ISO) or with unprecedented spatial resolution at their times (Voyager and Cassini), despite being at more modest spectral resolutions (R $<$ 3000) than recent ground-based observations. First results from the James Webb Space Telescope are currently being published and will be discussed further in this paper \citep{Hueso2023, RodriguezOvalle2024}.

Similarly, millimeter observatories have evolved drastically, starting with spatially unresolved millimetric observations with the Odin telescope \citep{Cavalie2008b, Cavalie2012, Benmahi2020}, and observations containing only a few resolution elements across the Jovian disk with the James Clerk Maxwell telescope (JCMT) and IRAM-30m telescope \citep{Lellouch1995, Lellouch1997, Moreno2003}. Large interferometric facilities such as the Atacama Large Millimeter Array (ALMA) and the Northern Extended Millimeter Array (NOEMA) now combine high spectral and spatial resolution observations of the gas giants \citep{Cavalie2021, Benmahi2022} resulting in an improved understanding of the different processes occurring in the low- and polar latitudes affected by the auroral activity.


Spacecraft missions, such as Pioneer, Voyager, Galileo, Cassini-Huygens and Juno, have provided valuable observations of Jupiter's upper atmosphere across a range of spectral windows, in particular on the night sides of the outer planets, inaccessible to ground and Earth orbiting telescopes \citep[e.g.][]{Bolton2017}. Unlike ground-based observations, spacecraft are able to observe the planet's nightside as well as the dayside and have the opportunity to obtain high spatial resolution measurements. The Juno mission has provided a particularly unique view of Jupiter due to its highly eccentric polar orbit; Jupiter's low obliquity (3.13$^{\circ}$) means that only limited views of Jupiter's poles can be obtained from Earth's vantage point, so the observations obtained by Juno have transformed our understanding of Jupiter's polar atmosphere \citep{Bolton2017}.

This paper summarizes the current knowledge of Jupiter's upper atmospheric chemical distributions and dynamics, with a focus on polar latitudes, as derived from ground- and space-based observations acquired over the last decades. Jupiter's polar regions host one of the most extreme planetary environments in the solar system. The magnetosphere-ionosphere interaction resulting from its powerful magnetic field, coupled with plasma-rich nearby magnetospheric environment broadly affect the high-latitude thermal structure and chemical distributions, as reviewed here. Auroral particle precipitation produces some of the brightest auroras, as observed across wavelengths ranging from X-ray to radio. Elevated stratospheric temperatures due to charged particle precipitation have been measured for decades using ground- and spaced-based observations, as well as the consequences of this coupling for the high-latitude chemical and dynamics. 

First, we review the measured chemical distributions while emphasizing the spatially resolved observations that include the high-latitude regions. Next we discuss the various existing models (photochemical models, ion-neutral chemical models, and aerosol formation models) that aim to explain these distributions. Third, we discuss observations constraining the magnetosphere-ionosphere coupling through measurements of upper atmospheric winds. Finally, remaining outstanding questions will be summarized and discussed.

\section{Stratospheric chemical distributions}
\label{s:chem}

\subsection{Hydrocarbons}
\label{ss:HC}

\subsubsection{Constraints from mid-IR observations}

One of the prime wavelength ranges to monitor hydrocarbon emissions is the mid-IR, as, e.g., observed earlier from the Voyager 1 and 2 missions, or using space telescopes such as ISO \citep{Encrenaz1996, Fouchet2000}. The mid-IR is an important wavelength region containing spectral signatures of many atmospheric species that contribute to regulate the stratospheric radiative balance \citep[e.g.][]{Roman2023}. The main drawback is the degeneracy between temperature and abundances of the observed species, as they produce similar effects on spectra, and both are physically plausible in auroral regions. Studies discussed in this section make different assumptions about temperature profiles. 


Measurements made with IRIS on Voyager 1 and 2 provided some of the first spatial information about the hydrocarbon distribution \citep{Hanel1979a, Hanel1979b}. These observations indicated a general mid-IR polar brightening, reflecting either an enhancement in stratospheric temperature and/or an increase in the hydrocarbon abundances \citep{Kim1985, Drossart1993}. The presence of these stratospheric regions with elevated thermal emissions (nicknamed \textit{hot spots}) was similarly detected using ground-based facilities \citep{Caldwell1980, Orton1991}, or prior Voyager data. By deriving a thermal profile in the polar region required to fit the CH$_4$ emissions, whose abundance was assumed to be constant across the planet, \cite{Kim1985} measured an abundance enhancement with respect to mid-latitudes. In the polar IR-bright region (averaging latitudes $>$ 48$^{\circ}$N and longitudes 120-240$^{\circ}$W), they derived abundance enhancement factors of $\sim$3 and $\sim$2 with respect to the mid-latitudes (averaging latitudes 30$^{\circ}$S-0$^{\circ}$, excluding longitudes 120-220$^{\circ}$W), for C$_2$H$_2$ and C$_2$H$_6$, respectively. They also derived abundance measurements of C$_2$H$_4$, C$_6$H$_6$, and CH$_3$C$_2$H in the northern polar region, though only upper limits for these molecular abundances were previously derived at mid-latitudes.

Ground-based observations using infrared heterodyne spectroscopy techniques allowed further investigation of the C$_2$H$_\mathrm{x}$ emissions in the polar regions of Jupiter. \citet{Livengood1993} and \citet{Kostiuk1993} were able to derive scaling factors of the C$_2$H$_6$ and C$_2$H$_4$ abundances needed to reproduce the observed spectra, assuming a temperature profile based on Voyager 2 IRIS data recorded several years prior. However, because of the degeneracy between temperature and hydrocarbon abundances (also illustrated by \citealt{Drossart1993}), these studies provided a range of acceptable temperature and hydrocarbon abundances within the auroral regions. Another challenging aspect of measuring spatial trends in polar hydrocarbon emissions using ground-based telescopes relates to the observation geometry. Because of Jupiter's small obliquity (3.13$^{\circ}$) combined with the rather small magnetic dipole tilt (10.2$^{\circ}$, according to the Juno prime mission model of \citet{Connerney2022}), the auroral regions are only observable at high-emission angles during a fraction of the Jovian rotation. Additionally, Jupiter's magnetic field is asymmetric, and the northern aurora extends more equatorward than the southern one, making the former one easier to observe from Earth.


IR measurements were similarly acquired using the Cassini/CIRS instrument during the December 2000 Jupiter flyby \citep{Kunde2004}. \cite{Nixon2007} retrieved the meridional variations of temperature, as well as the C$_2$H$_2$ and C$_2$H$_6$ abundances in the 70$^{\circ}$S-70$^{\circ}$N range. Temperatures were derived from spectral features of H$_2$ and CH$_4$. The continua at 600–670\,cm$^{-1}$ (14.9-16.6\,$\mu$m) and 760–800\,cm$^{-1}$ (12.5-13.2\,$\mu$m) are due to the broad H$_2$ S(1) line, and were used to retrieve the upper-tropospheric temperature, while assuming a constant H$_2$ gas abundance. The CH$_4$ $\nu_4$ band at 1225–1325\,cm$^{-1}$ (7.5-8.2\,$\mu$m) was used to retrieve the stratospheric temperatures. The C$_2$H$_2$ and C$_2$H$_6$ abundances were subsequently retrieved using their respective $\nu_5$ band at 670-760\,cm$^{-1}$ (13.2-14.9\,$\mu$m) and $\nu_9$ band at 800-850\,cm$^{-1}$ (11.8-12.5\,$\mu$m). The main contributions to the C$_2$H$_2$ and C$_2$H$_6$ emissions are located around 5\,mbar, with C$_2$H$_2$ contributions extending up to 0.1\,mbar and 10\,$\mu$bar.

At high-latitude, \cite{Nixon2007} derived the meridional distribution of C$_2$H$_2$ and C$_2$H$_6$ while zonally averaging the CIRS-acquired spectra. CIRS revealed an enhancement in the emission of several hydrocarbons around a large region located near the magnetic poles of Jupiter. Consequently, \cite{Nixon2007} excluded from their zonal average spectra recorded at 60$^{\circ}$S-70$^{\circ}$S and 330-90$^{\circ}$W, as well as 60$^{\circ}$N-70$^{\circ}$N and 150-210$^{\circ}$W, where such hot spots had been repeatedly identified. Note that throughout this document, all quoted latitudes are planetocentric. At 5\,mbar in the stratosphere, \cite{Nixon2007} found that the C$_2$H$_2$ and C$_2$H$_6$ meridional trends are mainly anticorrelated, with C$_2$H$_2$ decreasing with increasing latitude, while C$_2$H$_6$ peaks at $\pm$70$^{\circ}$. This result poses a dilemma as both hydrocarbons are by-products of the methane photolysis and therefore should follow similar spatial trends (see section \ref{ss:neutral_models} for a model/data comparison). The anti-correlated trends were then assumed to be caused by a difference in chemical lifetime between these two molecules.

\cite{Nixon2010} subsequently performed a detailed C$_2$H$_2$ and C$_2$H$_6$ spectral modeling of the IRIS measurements obtained during the Voyager 1 flyby of Jupiter in 1979. Using a similar retrieval method, they confirmed the finding from \cite{Nixon2007} while reanalyzing both datasets using an updated spectral database for ethane. Extending the analysis of the CIRS dataset, \cite{Sinclair2017} retrieved the temperature, C$_2$H$_2$, C$_2$H$_4$ and C$_2$H$_6$ distributions as a function of latitude and longitude, including in the auroral hot spot region. They tested the robustness of the temperature retrieval using several thermal a priori profiles with variable upper stratospheric temperatures. They found, using both Cassini/CIRS high ($\Delta\,\nu$ = 0.5 \,cm$^{-1}$) and medium ($\Delta\,\nu$ = 2.5 \,cm$^{-1}$) spectral setup the presence of localized hot spots at two different pressure levels ($\sim$10\,$\mu$bar and $\sim$1\,mbar) with elevated stratospheric temperatures on the order of 20\,K and 10\,K, respectively, though the exact magnitude was found to depend on the spectral setup of the instrument. The Voyager/IRIS measurements, on the other hand, did not support the existence of that double-peaked localized heating regions, which was attributed to a lack of sensitivity due to IRIS' lower spectral resolution.

\citet{Sinclair2017} further studied the enhancement in C$_2$H$_6$ and depletion in C$_2$H$_2$ at non-auroral high-latitudes. Above Jupiter's main UV-auroral ovals, they additionally showed a localized $\sim$15\% enhancement of C$_2$H$_2$ in the 1-5\,mbar range while also showing a depletion in C$_2$H$_6$, despite rather large error bars. \citet{Sinclair2017} showed a key figure (their Fig. 13) demonstrating that, at 70$^{\circ}$N, the combined amount of C$_2$H$_2$+C$_2$H$_6$ stayed constant as a function of longitude, while the local enhancement and depletion of these two hydrocarbons compensating for each other in the longitudinal region of the main oval. They suggested that C$_2$H$_2$ may be locally produced within the auroral region and converted into C$_2$H$_6$ outside of the oval.

\cite{Sinclair2019} extended the modeling of the Voyager/IRIS and Cassini/CIRS datasets by analyzing the emissions from further hydrocarbons: ethylene (C$_2$H$_4$) around 950\,cm$^{-1}$ (10.5\,$\mu$m), methylacetylene (CH$_3$C$_2$H) around 632\,cm$^{-1}$ (15.8\,$\mu$m), diacetylene (C$_4$H$_2$) around 628\,cm$^{-1}$ (15.9\,$\mu$m), and benzene (C$_6$H$_6$) around 674\,cm$^{-1}$ (14.8\,$\mu$m). In the auroral regions, they found that abundance enhancements of 6.4 and 9.6 in C$_2$H$_4$ and C$_6$H$_6$ with respect to photochemical model A of \cite{Moses2005a} were needed to reproduce the high-spectral resolution ($\Delta\,\nu$ = 0.5 \,cm$^{-1}$) CIRS spectra. Similarly, auroral enhancements of 1.6, 3.4 and 15 in the abundances of C$_2$H$_4$, CH$_3$C$_2$H, and C$_6$H$_6$, respectively, were needed to reproduce the medium resolution CIRS spectra ($\Delta\,\nu$ = 2.5 \,cm$^{-1}$). Only an upper limit of C$_4$H$_2$ was found in the auroral and non-auroral regions.

\begin{figure}[!h]
 \begin{center}
    \includegraphics[width=0.97\textwidth]{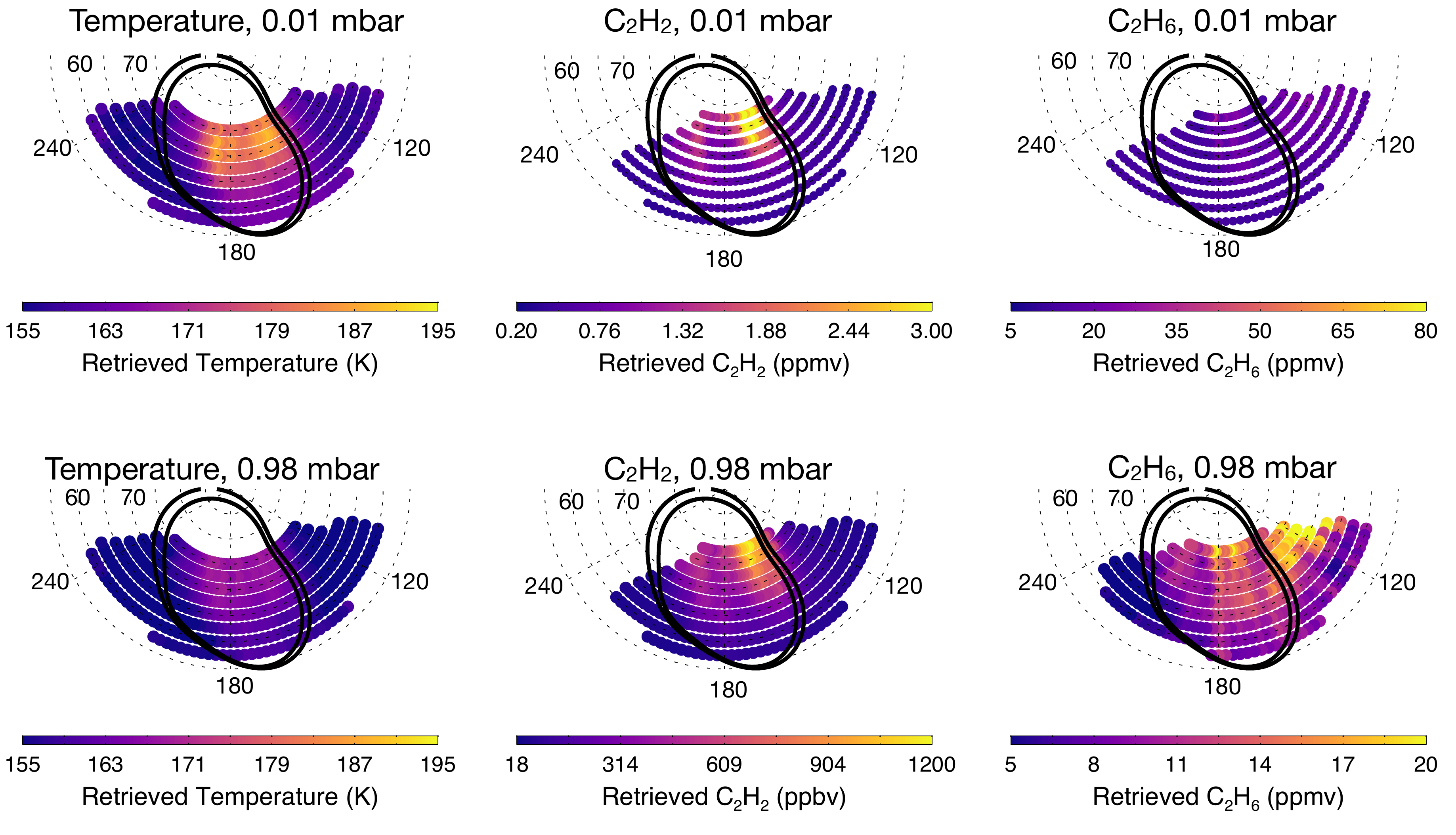}
  \end{center}
  \caption{Northern polar projection of the retrieved temperature (left), C$_2$H$_2$ (center) and C$_2$H$_6$ (right) abundances at 0.01\,mbar and 0.98\,mbar on 2017 March 19 from Gemini-N/TEXES \citep{Sinclair2023}. The contracted and expanded reference UV auroral oval derived from HST observations are displayed in solid black line \citep{Bonfond2012a}.}
  \label{fig:Sinclair2023a}
\end{figure}

\begin{figure}[!h]
 \begin{center}
    \includegraphics[width=0.97\textwidth]{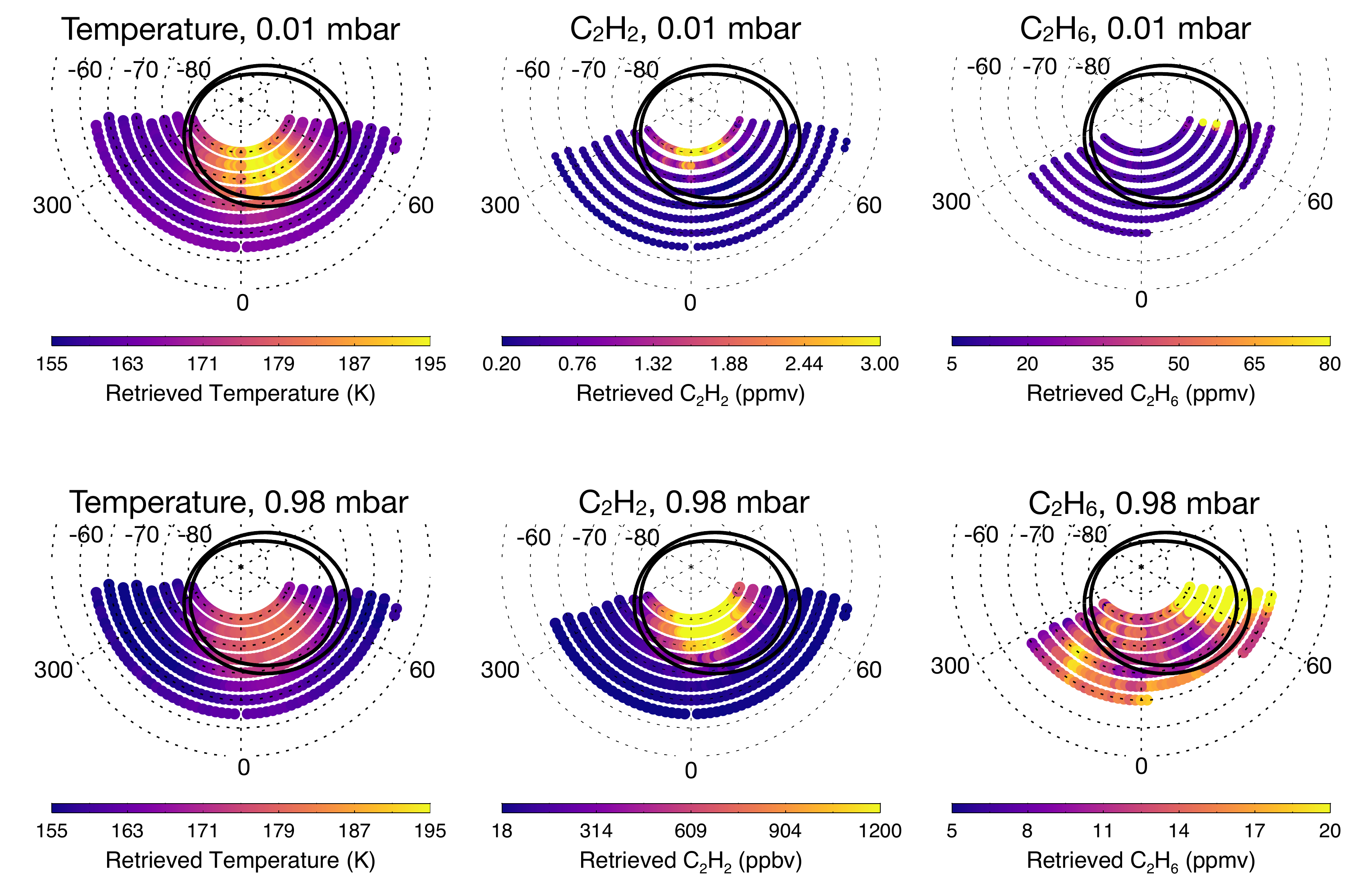}
  \end{center}
  \caption{Same as Figure \ref{fig:Sinclair2023a} for the southern hemisphere, from \citet{Sinclair2023}.}
  \label{fig:Sinclair2023b}
\end{figure}

Some of these findings triggered subsequent ground-based observing campaigns at higher spectral resolving power using the Texas Echelon Cross Echelle Spectrograph (TEXES) instrument \citep{Lacy2002}, mounted on the NASA Infrared Telescope Facility (IRTF) or the Gemini-North telescope \citep{Sinclair2020, Sinclair2023}. TEXES allowed the derivation of the meridional distribution of acetylene and ethane across Jupiter, confirming the general equator-to-pole decrease/increase in the C$_2$H$_2$/C$_2$H$_6$ abundances, respectively found by Cassini-CIRS \citep{Fletcher2016, Melin2018}. In the polar regions, despite the slightly less favorable viewing geometry than with Voyager or Cassini, TEXES observations permitted the derivation of the high-latitude temperature, C$_2$H$_2$, C$_2$H$_4$, and C$_2$H$_6$ abundance profiles. Figures \ref{fig:Sinclair2023a} and \ref{fig:Sinclair2023b} provide a snapshot at the Gemini-N/TEXES observations published by \cite{Sinclair2023}. They confirmed (i) the presence of the localized heating in the auroral region at two discrete pressure regions of 1\,mbar and 10\,$\mu$bar, (ii) the localized enhancement of C$_2$H$_2$, C$_2$H$_4$ within the auroral region, and (iii) enhancement in the C$_2$H$_6$ abundance over a large longitudinal range at high-latitude in the north polar region. \citet{Sinclair2023} also suggested that the local enhancements in temperature and hydrocarbons may be even more localized within the auroral region following a solar wind compression event.

The James Webb Space Telescope has recently been providing new mid-IR observations of Jupiter, using the medium resolution spectroscopy modes (MIRI-MRS), as part of the Early Release Science ERS 1373 program. \citet{RodriguezOvalle2024} retrieved the thermal structure and hydrocarbon abundances within the southern polar region. They confirmed the earlier findings from \citet{Sinclair2017} on the double-peaked thermal structure, with the two hotpots located at 1\,mbar and 0.01\,mbar. JWST/MIRI experiences saturation when observing bright bodies, and \citet{RodriguezOvalle2024} developed a specific reduction procedure to desaturate the data and enable the derivation of the C$_2$H$_2$ and C$_2$H$_6$ abundances. They measured enhancements by factors of $\sim$5 and $\sim$7 in the C$_2$H$_2$ and C$_2$H$_6$ abundances at 7\,mbar and 3\,mbar, respectively, within the southern polar auroral region (75$^{\circ}$S), compared to the region outside the auroral region (60$^{\circ}$S).

Tables \ref{tab:C2H2_tab}, \ref{tab:C2H4_tab}, \ref{tab:C2H6_tab}, and \ref{tab:CxHy_tab} summarize the spatially-resolved hydrocarbon measurements including the high-latitude regions. These tables only summarize the observations that led to the derivation of a spatial trend in the hydrocarbon abundances at high-latitudes: either meridional trends, or zonal enhancement at polar auroral latitudes. For a list of prior globally averaged observations, see e.g., \citet{Moses2005a} and \citet{Sada1998}. For the corresponding coverage of each of these observing campaigns in terms of latitudinal range observed and pressure levels probed, the reader is referred to Appendix \ref{sec:appendixA}. Prior studies have derived upper limits for diacetylene (C$_4$H$_2$), and propane (C$_3$H$_8$), but these are not listed since no spatial trends were inferred. Note that the acetylene contribution function extents higher in the stratosphere, and also provides abundance information around $\sim$\,0.1\,mbar, though not listed here for the sake of brevity and consistency with the other hydrocarbons.


\begin{table}[ht]
\begin{center}   
\footnotesize    
\begin{tabular}{p{1.5cm} p{2.5cm} p{7cm} p{0.35cm} } 
 \hline
Instrument & Coverage & Trend & Ref  \\
 \hline

Voyager/IRIS & 57$^{\circ}$S-56$^{\circ}$N (auroral regions excluded) & Constant with latitude at 0.1-7\,mbar & [1] \vspace{0.35cm} \\ 

Voyager/IRIS & 67$^{\circ}$N-72$^{\circ}$N & Zonal enhancement by a factor of 9 at 5\,mbar in auroral region & [2] \vspace{0.35cm} \\
           
Cassini/CIRS & 66$^{\circ}$S-71$^{\circ}$N (auroral regions excluded) & Decreases toward poles by a factor of 4 at 5\,mbar, north-south asymmetry (maximum at 20$^{\circ}$N) & [3] \vspace{0.35cm} \\
           
Cassini/CIRS & 65$^{\circ}$N-85$^{\circ}$N & Zonal enhancement by a factor of 6 at 5\,mbar in auroral region & [2]   \\
 & 85$^{\circ}$S-70$^{\circ}$S & Zonal enhancement by a factor of 3 at 5\,mbar in auroral region &  \vspace{0.35cm}  \\
           
IRTF/TEXES & 70$^{\circ}$S-70$^{\circ}$N  & Decreases toward poles by a factor of 2 at 5\,mbar, north-south asymmetry (maximum at 20$^{\circ}$N) & [4] \vspace{0.35cm} \\
           
IRTF/TEXES & 70$^{\circ}$S-70$^{\circ}$N & Decreases toward poles by factors of 2–3 at 1\,mbar, north-south asymmetry decreases over 4 yr of observations & [5]  \vspace{0.35cm} \\
           
IRTF/TEXES & 45$^{\circ}$N-75$^{\circ}$N & Zonal enhancement by a factor of 2 at 5\,mbar in auroral region & [6] \\
  & 80$^{\circ}$S-45$^{\circ}$S & No significant zonal trend in auroral region &   \vspace{0.35cm} \\
           
Juno/UVS & 75$^{\circ}$S-75$^{\circ}$N (auroral regions excluded) & Decreases toward poles by a factor of 2-4 at pressures less than 5–50\,mbar & [7]  \vspace{0.35cm} \\
           
Juno/UVS & Full coverage & Auroral enhancement by a factor of 2 (north) and 3.5 (south) & [8]  \vspace{0.35cm} \\
           
Gemini-N/TEXES & 52$^{\circ}$N-70$^{\circ}$N & Auroral enhancement by a factor of 10-23 at 1\,mbar & [9]  \\
  & 80$^{\circ}$S-62$^{\circ}$S & Auroral enhancement by a factor of $\sim$40 at 1\,mbar & \vspace{0.35cm} \\  
           
JWST/MIRI & 55$^{\circ}$S-80$^{\circ}$S & Auroral enhancement by a factor of 5 at 7\,mbar & [10]  \\
\hline 
\end{tabular}
\caption{Summary of Jovian acetylene (C$_2$H$_2$) measurements. 
[1] \citet{Nixon2010} ;
[2] \citet{Sinclair2017} ;
[3] \citet{Nixon2007} ;
[4] \citet{Fletcher2016} ;
[5] \citet{Melin2018} ;
[6] \citet{Sinclair2018} ;
[7] \citet{Giles2021b} ;
[8] \citet{Giles2023} ;
[9] \citet{Sinclair2023} ;
[10] \citet{RodriguezOvalle2024}}
\label{tab:C2H2_tab}
\end{center}
\end{table}

\begin{table}[ht]
\begin{center}   
\footnotesize    
\begin{tabular}{p{1.5cm} p{2.5cm} p{7cm} p{0.35cm} } 
 \hline
Instrument & Coverage & Trend & Ref  \\
 \hline
 Voyager/IRIS & 60$^{\circ}$N-80$^{\circ}$N & Scaling factor of 1.6 in auroral region compared to models & [1] \vspace{0.35cm} \\
 
 Cassini/CIRS & 65$^{\circ}$N-85$^{\circ}$N & Scaling factor of 7.2 compared to models & [1] \\
              & 85$^{\circ}$S-70$^{\circ}$S & Scaling factor of 7.2 compared to models &  \vspace{0.35cm} \\
              
 IRTF/TEXES & 45$^{\circ}$N-75$^{\circ}$N & Zonal enhancement by a factor of 10 at 5\,$\mu$bar in auroral region & [2] \\
            & 80$^{\circ}$S-45$^{\circ}$S & Zonal enhancement by a factor of 1.7 at 5\,$\mu$bar in auroral region & \vspace{0.35cm} \\
            
 Gemini-N/TEXES & 52$^{\circ}$N-70$^{\circ}$N & Zonal enhancement by a factor of 13 at 1\,$\mu$bar in auroral region & [3] \\
                & 80$^{\circ}$S-62$^{\circ}$S & Zonal enhancement by a factor of 18 at 1\,$\mu$bar in auroral region & \\
\hline 
\end{tabular}
\caption{Summary of Jovian ethylene (C$_2$H$_4$) measurements.
[1] \citet{Sinclair2019} ;
[2] \citet{Sinclair2018} ;
[3] \citet{Sinclair2023}
}
\label{tab:C2H4_tab}
\end{center}
\end{table}

\begin{table}[ht]
\begin{center}   
\footnotesize    
\begin{tabular}{p{1.5cm} p{2.5cm} p{7cm} p{0.35cm} } 
 \hline
Instrument & Coverage & Trend & Ref  \\
 \hline
Voyager/IRIS & 57$^{\circ}$S-56$^{\circ}$N (auroral regions excluded) & Increases toward poles by a factor of 2-3 at 5\,mbar & [1] \vspace{0.35cm} \\ 

Voyager/IRIS & 67$^{\circ}$N-72$^{\circ}$N & Zonal decrease by a factor of $\sim$2 at 5\,mbar in auroral region & [2] \vspace{0.35cm} \\
           
Cassini/CIRS & 66$^{\circ}$S-71$^{\circ}$N (auroral regions excluded) & Increases toward poles by a factor of 1.6 at 5\,mbar & [3] \vspace{0.35cm} \\
           
Cassini/CIRS & 65$^{\circ}$N-85$^{\circ}$N & Zonal decrease by a factor of $<$1.4 at 5\,mbar in auroral region & [2]   \\
 & 85$^{\circ}$S-70$^{\circ}$S & Zonal decrease by a factor of 1.5 at 5\,mbar in auroral region &  \vspace{0.35cm}  \\
           
IRTF/TEXES & 70$^{\circ}$S-70$^{\circ}$N & Increases toward poles by a factor of 1.5-2 at 5\,mbar &  [4] \vspace{0.35cm} \\
           
IRTF/TEXES & 70$^{\circ}$S-70$^{\circ}$N & Increases toward poles by factors of 1.8 at 1\,mbar & [5] \vspace{0.35cm} \\
           
IRTF/TEXES & 45$^{\circ}$N-75$^{\circ}$N & Lower retrieved concentration within auroral region at $<$1\,mbar, not statistically significant & [6] \\
  & 80$^{\circ}$S-45$^{\circ}$S & Lower retrieved concentration within auroral region at $<$1\,mbar, not statistically significant &   \vspace{0.35cm} \\
  
Gemini-N/TEXES & 52$^{\circ}$N-70$^{\circ}$N & Zonal enhancement by a factor of 2-3 at 5\,mbar in auroral region & [7]  \\
  & 80$^{\circ}$S-62$^{\circ}$S & Zonal enhancement by a factor of 1.8 at 5\,mbar in auroral region &  \vspace{0.35cm} \\
  
JWST/MIRI & 55$^{\circ}$S-80$^{\circ}$S & Auroral enhancement by a factor of 7 at 3\,mbar & [8]  \\
\hline 
\end{tabular}
\caption{Summary of Jovian ethane (C$_2$H$_6$) measurements. 
[1] \citet{Nixon2010} ;
[2] \citet{Sinclair2017} ;
[3] \citet{Nixon2007} ;
[4] \citet{Fletcher2016} ;
[5] \citet{Melin2018} ;
[6] \citet{Sinclair2018} ;
[7] \citet{Sinclair2023} ;
[8] \citet{RodriguezOvalle2024}
}
\label{tab:C2H6_tab}
\end{center}
\end{table}

\begin{table}[ht]
\begin{center}   
\footnotesize    
\begin{tabular}{p{1.5cm} p{2.5cm} p{7cm} p{0.35cm} } 
 \hline
Instrument & Coverage & Trend & Ref  \\
 \hline
 \hline
 CH$_3$C$_2$H  &  &  &   \\
 \hline
 Voyager/IRIS & 60$^{\circ}$N-80$^{\circ}$N & Scaling factor of 3.4 in auroral region compared to models & [1] \vspace{0.35cm} \\
 Cassini/CIRS & 65$^{\circ}$N-85$^{\circ}$N & Scaling factor of $<$9.6 compared to models & [1] \\
              & 85$^{\circ}$S-70$^{\circ}$S & Scaling factor of 3.4 compared to models &   \\
 \hline 
 \hline
 C$_6$H$_6$  &  &  &  \\
 \hline
  Voyager/IRIS & 60$^{\circ}$N-80$^{\circ}$N & Scaling factor of 15.6 in auroral region compared to models & [1] \vspace{0.35cm} \\
 Cassini/CIRS & 65$^{\circ}$N-85$^{\circ}$N & Scaling factor of 10.6 compared to models & [1] \\
              & 85$^{\circ}$S-70$^{\circ}$S & Scaling factor of 10.4 compared to models &   \\
 JWST/MIRI & 55$^{\circ}$S-80$^{\circ}$S & Auroral enhancement by a factor of 9.7 at 0.5\,mbar & [2]  \\
 \hline
\end{tabular}
\caption{Summary of Jovian methylacetylene (CH$_3$C$_2$H) and benzene (C$_6$H$_6$) measurements.
[1] \citet{Sinclair2019}
[2] \citet{RodriguezOvalle2024}
}
\label{tab:CxHy_tab}
\end{center}
\end{table}

\subsubsection{Observational constraints on the polar upper-atmospheric mixing processes}

Hydrocarbon mid-IR observations, along with UV-occultation techniques, provide indirect information about the mixing processes controlling the upper atmospheric vertical structure. In the upper stratosphere, the strength of the molecular diffusion increases more rapidly with decreasing pressures than does the eddy diffusion. The species therefore experience a diffusive separation, as their vertical structures become controlled by their own molecular weight above their respective homopause \citep[e.g.][]{Gladstone1996}. Methane, the third most abundant molecule in the lower stratosphere, is well-mixed within the lower to middle stratosphere and its abundance decreases due to molecular diffusion near the homopause and above. The hydrocarbon absorption in specific parts of the far-UV spectrum, combined with the sharp decrease in the hydrocarbon concentration above the homopause, allows the derivation of their concentrations using stellar- and solar-occultation techniques.

Several studies discussing the various UV stellar occultations available tend to disagree regarding the derived methane mole fraction at the homopause (see \citet{Moses2005} for a discussion regarding the Voyager-UVS stellar occultation analysis of \citet{Festou1981} and \citet{Yelle1996}). The main drawback comes from the limited knowledge of the background atmospheric structure (temperature and wind structures, planetary shape, vertical variation of the mean molecular mass) at the times and locations these occultations are performed. Note that, on Saturn, this situation is slightly better thanks to the simultaneous CIRS and UVIS measurements, allowing the reconstruction with higher precision the atmospheric structure at the location of the stellar occultation \citep{Koskinen2018, Brown2024}.

Using New-Horizons Alice occultation measurements of $\chi$-Ophiuchus, \citet{Greathouse2010} derived additional values for the eddy diffusion ($K_{zz}$) coefficient near the methane homopause at two non-auroral latitudes. While the derived value (3.4$^{+9.0}_{-2.8}$ $\times$ 10$^{6}$ cm$^2$s$^{-1}$) agrees with the previous Voyager-based determination \citep{Festou1981, Yelle1996}, the corresponding pressure level of the methane homopause differed from these previous determinations. This possibly suggests a change in the strength of the atmospheric mixing.

Despite uncertainty in the derivation of the methane homopause, the range of $K_{zz}$ provided at non-auroral latitude does not seem to agree with those derived in the polar region. Using Cassini-UVIS measurements, \citet{Parkinson2006} could not reproduce the Helium 58.4\,nm airglow intensities considering the known eddy diffusion coefficient magnitude derived at lower latitudes. This high-latitude non-solar sources of Helium emission are suspected to be caused by (i) enhanced electron impact, (ii) enhanced vertical mixing, or a combination of both. \citet{Parkinson2006} ruled out (i) using results from the \citet{Grodent2001} polar atmosphere model, suggesting that the eddy diffusion coefficient in the polar mesosphere is enhanced by a factor of 4-10 compared to its lower latitude values.

Using mid-IR infrared observations recorded from IRTF/TEXES, \citet{Sinclair2020} analyzed methane and methyl radical (CH$_3$) emission spectra measured to further constrain the high-altitude vertical structure in the polar region. Methyl is one of the main methane photolysis by-products, that forms ethane through methyl-methyl recombination. Its abundance is predicted to peak around the methane homopause and is highly dependent on the eddy diffusion coefficient. Initially detected by Cassini-CIRS on Jupiter \citep{Kunde2004}, its emission is optically thin and depends on the upper stratospheric temperature.

\citet{Sinclair2020} tested several photochemical model profiles from \citet{Moses2017}, considering various assumptions for the eddy diffusion coefficient at pressures less than 10\,$\mu$bars. Depending on the different model outputs, they produced synthetic spectra for these three molecules by allowing the temperature profile to vary. \citet{Sinclair2020} compared the synthetic emission spectra to the observed ones calculating a goodness-of-fit metric, and concluded that the methane homopause at high-northern latitudes is located at higher altitude. At 62$^{\circ}$N, the methane homopause inside the main oval seems located at 461$^{+147}_{-39}$\,km above the 1-bar level, i.e. $\sim$129\,km higher than elsewhere (away from the main oval or at lower latitudes). \citet{RodriguezOvalle2023EGU} derived the homopause altitude in the southern polar region from JWST using a similar technique with MIRI observations, suggesting that the methane homopause is located at $\sim$590$^{+17}_{-56}$\,km above the 1\,bar level within the southern main oval. These results seem to disagree with \citet{Kim2017}, who found the methane vertical profile in the north polar region relatively similar to the one derived in the equatorial region. This discrepancy is possibly due to the assumed vibrational relaxation rates of methane between these studies \citep[see, e.g., the discussion section of][]{Sinclair2020}, and/or the temporal variability in the studied emissions potentially linked with solar forcing. Indeed, the 8\,$\mu$m CH$_4$ emission was previously shown to display daily variability, possibly triggered by solar-wind compressions \citep{Sinclair2019}.


\subsubsection{Constraints from reflected UV observations}

Ultraviolet spectrum of Jupiter in the far-UV carries the absorption signature of several hydrocarbons. Using Cassini-UVIS, \cite{Melin2020} initially derived latitudinal trends for the acetylene and ethane abundance around the low- to mid-latitudes, up to $\sim$40$^{\circ}$. The Ultraviolet Spectrograph (UVS) on Juno similarly provides information on hydrocarbons, as reflected sunlight captured by UVS is partially absorbed by hydrocarbons and haze. The UVS instrument is a photon-counting imaging-spectrograph operating in the 68-210\,nm range \citep{Gladstone2017_SSR, Davis2011, Greathouse2013}. While the primary goal of UVS is to capture the Jovian UV aurora and provide context to Juno's in-situ instruments, it also captures a large amount of non-auroral UV Jovian emissions \citep{Giles2020, Giles2021a} and stellar observations \citep{Hue2019a, Hue2021b}, due to the spin-stabilized nature of the spacecraft.

\citet{Giles2021b} used Juno-UVS reflected sunlight observations to retrieve the mean C$_2$H$_2$ abundance as a function of latitude, excluding the auroral regions. They found a poleward decrease in its abundance by a factor of 2-4 at pressures less than 5-50\,mbar, consistent with prior non-auroral mid-IR observations \citep{Nixon2007, Fletcher2016, Melin2018}. \citet{Giles2023} then extended this work by analyzing Juno-UVS reflected sunlight maps of Jupiter's poles, including the auroral regions. They compared the mean reflectance at 175-190 nm, where there is absorption from both C$_2$H$_2$ and C$_6$H$_6$, with the mean reflectance at 190-205 nm, where the absorption from these hydrocarbons is much weaker. Within the main auroral emission, there is a significant enhancement in the 175-190 nm absorption relative to the non-auroral longitudes. This could either be due to an enhancement in the C$_2$H$_2$ abundance alone, or due to an enhancement in both C$_2$H$_2$ and C$_6$H$_6$; variations in C$_6$H$_6$ alone cannot reproduce the observations. If the observed variability is entirely due to C$_2$H$_2$, \citet{Giles2023} conclude that the C$_2$H$_2$ abundance within the auroral oval must be a factor of 3 larger than at similar latitudes but non-auroral longitudes.

\begin{figure}[!h]
 \begin{center}
    \includegraphics[width=0.8\textwidth]{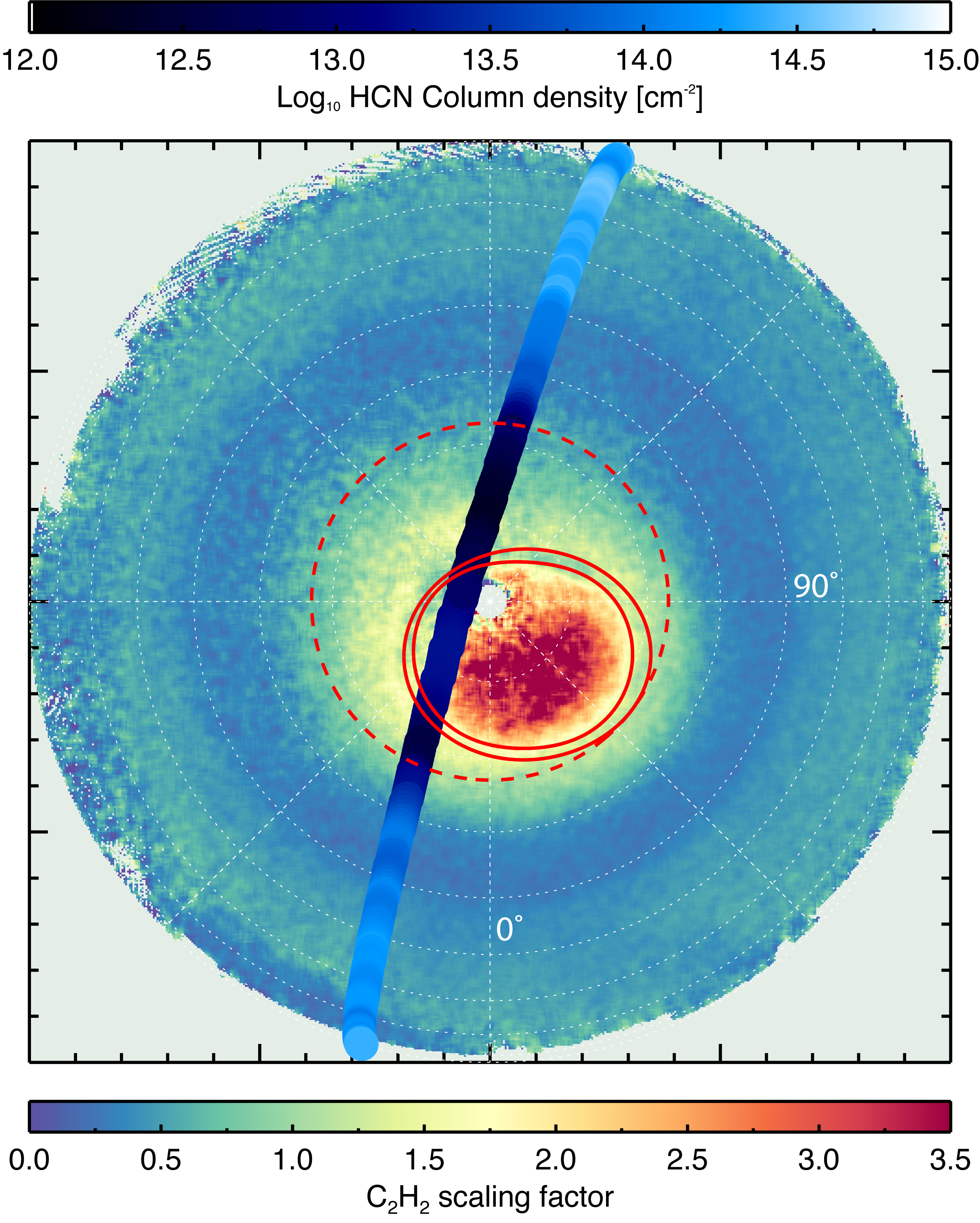}
  \end{center}
  \caption{Southern polar projection of the C$_2$H$_2$ abundance derived by \citet{Giles2023} using Juno-UVS reflected sunlight observations. The abundance is expressed as a scaling factor relative to the nominal vertical profile from \citet{Moses2005a}. This is compared with the ALMA observations of \citet{Cavalie2023b} who derived the HCN column density at Jupiter's limb (see section \ref{ss:Cyanide}). The contracted and expanded reference ovals derived from HST UV-observations are displayed in solid red/blue lines \citep{Bonfond2012a}. The lowest parallel reached by the expanded southern reference oval (67$^{\circ}$) is displayed as a dashed circle.}
  \label{fig:Giles2023_vs_Cavalie2023}
\end{figure}

Figure \ref{fig:Giles2023_vs_Cavalie2023} compares the C$_2$H$_2$ abundance derived in the southern hemisphere with the HCN column density measurements from ALMA (see section \ref{ss:Cyanide}). The contracted and expanded reference ovals displayed are derived from HST UV-observations \citep{Bonfond2012a}. The figure also shows the lowest parallel reached by the expanded southern reference oval. Because the aurora is rotating with the planet around the spin axis rather than the magnetic axis, which is significantly tilted with respect to the former, this boundary limits the polar region where the auroral emissions (or auroral-induced chemistry by-products) are most likely to be found.



\subsubsection{Summary on hydrocarbons}

All these observations demonstrate that hydrocarbons are locally affected around the polar auroral regions. Mid-IR observations suggest that there is an enhancement in the amount of C$_2$H$_2$, C$_2$H$_4$, CH$_3$C$_2$H, and C$_6$H$_6$ at the mbar and sub-mbar levels of the polar auroral regions, i.e., poleward of the main oval. The C$_2$H$_2$ auroral enhancement finding was reinforced by a similar conclusion reached using a different wavelength (far-UV), though sensitive to slightly higher pressure level (5–50 mbar). The picture regarding C$_2$H$_6$ is less clear. Although earlier Voyager/IRIS and Cassini/CIRS measurements suggest that there is a localized depletion of C$_2$H$_6$ at the mbar level in the polar auroral regions, more recent ground-based observations using IRTF/TEXES, Gemini-N/TEXES, or JWST/MIRI show either no statistically significant enhancement of the C$_2$H$_6$ abundance, or an increase by a factor of $<$7 in the auroral regions. Zonally averaged spectra excluding the longitudinal sector containing the aurora however seem to suggest that C$_2$H$_6$ is generally enhanced in the polar region compared to the low-to-mid latitudes, as suggested by the Cassini/CIRS measurements retrieved by \citet{Nixon2007, Nixon2010}.

\subsection{Oxygen species}
\label{ss:Oxy_species}

The main oxygen species detected in Jupiter's stratosphere are carbon monoxide (CO), water (H$_2$O), and carbon dioxide (CO$_2$), ranked according to decreasing abundance. Their stratospheric distributions may result from both internal and external sources, and depend on a balance between vertical and horizontal transport (diffusion and wind), photochemical production/loss processes. Various external sources may supply oxygen species in Jupiter's atmosphere: interplanetary dust particle (IDPs, see \citealt{Prather1978}), interaction with icy moons and rings \citep{Strobel1979}, and cometary impacts \citep{Lellouch1995}. Oxygen species can be sourced from the planet's deep interior through upward diffusion of species quenched from deep tropospheric thermochemical equilibrium \citep{Prinn1977, Fegley1988, Yung1988, Fegley1994, Bezard2002, Visscher2010, Cavalie2023a}, though water condenses at the tropopause cold trap. Deciphering the origin of the detected species is achieved through observations spatially resolving these planets \citep[e.g.][]{Lellouch2006, Cavalie2013, Cavalie2023b}, combined with photochemical and thermochemical modeling \citep{Bezard2002, Lellouch2002}.

\subsubsection{Carbon monoxide}
The Jovian stratospheric CO has both an internal and external origin \citep{Bezard2002}. The internal source results from a $\sim$solar H$_2$O deep abundance \citep{Li2020_Jupiter, Cavalie2023a}. In addition, CO was delivered in large quantities by the Shoemaker-Levy 9 (SL9) comet fragments \citep{Lellouch1997}, which impacted Jupiter at several longitudes and around planetocentric latitude of 44$^{\circ}$S in July 1994 \citep{Hammel1995}. SL9 deposited CO at pressure levels of $\sim$0.1\,mbar and above. Stratospheric CO appears to have a secondary external source, which was first believed to be the result of more ancient cometary impacts \citep{Bezard2002}, but more likely results from IDP \citep{Moses2017}. CO is also produced from CO$_2$ photolysis, and recycled back into CO$_2$ through reaction with OH. However, photochemical modeling suggests that CO can be considered as photochemically stable \citep{Moses1996}. Temporal evolution of its spatial distribution several years following the SL9 impact allowed the derivation of the total mass of CO brought into the atmosphere as well as the vertical and horizontal diffusion coefficients \citep{Bezard2002, Moreno2003}.

About 22.5 years after the SL9 impact, ALMA measured the CO abundance at Jupiter's limb \citep{Cavalie2023b}. Spectral inversion indicated a rather latitudinally uniform CO column density of 1.86\,$\pm$\,0.52\,$\times$\,10$^{16}$\,cm$^{-2}$ (see fig. \ref{fig:Oxy}, top panel). This corresponds to a loss factor of 0.9\,$\pm$\,0.3 since the measurements of \citet{Moreno2003} recorded between 1995 and 1998 using the JCMT 15m and IRAM 30m telescopes.

\subsubsection{Water}
Condensation generally prevents H$_2$O from diffusing from the troposphere to the stratosphere of Jupiter, while spurious storm outbreaks may periodically overshoot the tropopause cold trap \citep{SanchezLavega2008}. Water was directly detected in the atmosphere following the SL9 impact \citep{Bjoraker1996, Encrenaz1997}. Measurements 3 years after the impact using ISO indicated that stratospheric water was restricted to pressure levels p $<$ 0.5 $\pm$ 0.2\,mbar. This was interpreted as a strong indication that the stratospheric water resulted directly from SL9 and subsequent chemistry \citep{Lellouch2002}.

Observations from the Submillimeter Wave Astronomy Satellite (SWAS) as well as early Odin disk-averaged observations lacked the sensitivity to further constrain the external source of H$_2$O \citep{Bergin2000, Cavalie2008b}. However, the Odin space telescope provided long-term monitoring of the globally-averaged stratospheric water, which tended to confirm the cometary origin of H$_2$O \citep{Cavalie2012, Benmahi2020}. 

The strongest evidence for the origin of water in Jupiter's stratosphere came from Herschel HIFI (Heterodyne Instrument for the Far Infrared) and PACS (Photodetector Array Camera and Spectrometer) observations. Disk-resolved mapping of the water emission at 66.4\,$\mu$m showed a factor of 2-3 decrease in the water column density from the southern to the northern hemisphere \citep{Cavalie2013}. Meridional cross-sections of the column density derived from the Herschel/PACS observations are shown on Figure \ref{fig:Oxy}, and compared with other oxygen species. The peak in the water column density around 15$^{\circ}$S likely results from a localized warmer temperature observed at this latitude that was not accounted for in the spectral modeling.

Over the years, the water emission line was expected to become fainter, as water diffused horizontally across Jupiter's disk, and more broadly, as the SL9-related stratospheric water diffused down to higher pressure levels. \cite{Benmahi2020} showed that the line-to-continuum ratio indeed decreased by $\sim$40\,\% between 2002 and 2019, and that the H$_2$O cutoff level migrated downwards from 0.2\,mbar in 2002 to 5\,mbar in 2019. Photochemical models can reproduce such a decrease provided that the efficiency of the vertical eddy diffusion coefficient increases in this pressure range. This in turns degrades the model agreement with the observed hydrocarbons, suggesting that an additional process might be responsible for the water loss. \citet{Benmahi2020} suggested that auroral chemistry may be the reason for this additional loss through ion-neutral chemistry triggered by electron precipitation.

\subsubsection{Carbon Dioxide}
The situation for CO$_2$ is not as straightforward since it can be created from H$_2$O photolysis and subsequent reaction with CO, which has a time-variable spatial distribution. Disk-resolved ISO observations initially indicated latitudinal variations, with a factor of $\sim$7 decrease in its column density from southern mid-latitudes to northern mid-latitudes \citep{Lellouch2002}. The ISO spatial resolution was however modest, only providing a couple of resolution elements across Jupiter.

The Cassini flyby of Jupiter provided higher spatial resolution observations, showing a strong CO$_2$ column density peak around the south pole \citep{Kunde2004, Lellouch2006}. Reconciling both the meridional distribution of CO$_2$ with the known post-SL9 distributions of H$_2$O and CO was a challenge. While minor CO$_2$ polar excess could be interpreted as a combination of the water photolysis by-product (OH) reacting with carbon monoxide, the observed magnitude of the CO$_2$ polar excess requires a different explanation. Magnetosphere-ionosphere coupling provides a means to locally supply ions originating from Io into the polar region, though the required quantity of oxygen ions to explain the CO$_2$ spatial distribution was unrealistically large \citep{Lellouch2006}. Finally, the CO$_2$ polar excess was explained using a combination of meridional diffusion combined with a poleward advection (discussed in section \ref{s:models}), though it has remained not a fully satisfactory solution given the transport constraints resulting from the HCN distribution measured simultaneously (see next section).

\begin{figure}[!h]
 \begin{center}
    \includegraphics[width=0.9\textwidth]{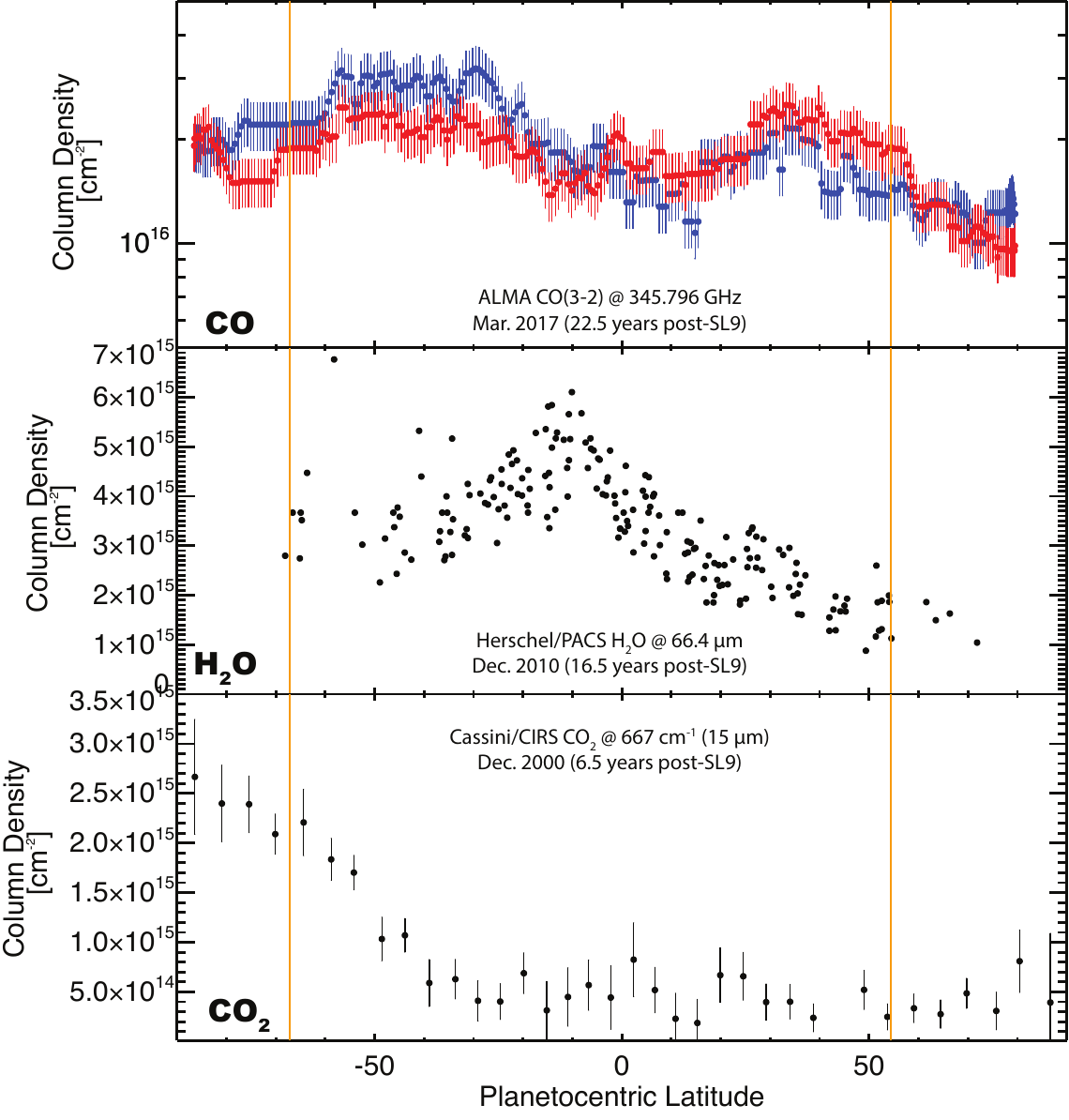}
  \end{center}
  \caption{Top panel: meridional CO column density derived from the CO (3-2) line observed by ALMA on 22 March 2017 at Jupiter eastern (red) and western (blue) limb. Adapted from \cite{Cavalie2023b}. Middle panel: meridional H$_2$O column density derived from Herschel/PACS observations at 66.4\,$\mu$m on 15 December 2010. Adapted from \cite{Cavalie2013}. Bottom panel: meridional CO$_2$ column density derived from Cassini/CIRS observations at 667 cm$^{-1}$ on December 2000. Adapted from \cite{Lellouch2006}.}
  \label{fig:Oxy}
\end{figure}

The CO, H$_2$O, and CO$_2$ meridional column densities have been measured at different post-SL9 times, which makes it difficult to find a unique interpretation. The measurements are summarized on Figure \ref{fig:Oxy}. Over-plotted in orange are the delimitation of Jupiter's auroral regions, defined here as the lowest parallels reached by the expanded reference ovals \citep{Bonfond2012a}. This corresponds to 54$^{\circ}$N and 67$^{\circ}$S.

\subsection{Nitrogen species}
\label{ss:Cyanide}

The only N-bearing species detected in Jupiter's stratosphere is HCN, while NH$_3$ is detected at pressures greater that $\sim$100\,mbar. Observations prior to the SL9 impact debated whether HCN was already present in the stratosphere in measurable quantities \citep{Tokunaga1981, Bezard1995}, and \citet{Bezard1995} only derived a mole fraction upper limit of 8$\times$10$^{-10}$ at p$<$55\,mbar. HCN was later observed in large quantities in Jupiter's stratosphere in the aftermath of the SL9 impact, with a total mass of HCN of 6$\times$10$^{11}$\,g on one well-studied impact fragment \citep{Marten1995}. 

While most HCN directly derives from impact-induced shock-chemistry \citep{Zahnle1996}, additional post-impact production resulted from the photolysis of NH$_3$ transported from the troposphere to the stratosphere during the impacts \citep{Griffith1997, Moses1995}. Temporal changes of its distribution during the years following the impact are mostly related to mixing processes (e.g., vertical and horizontal diffusions). This makes HCN an important species for tracing the Jovian dynamics in a region otherwise devoid of obvious tracers such as clouds. Monitoring its distribution up to 4 years after the impacts provided valuable information regarding the efficiency of the horizontal mixing \citep{Moreno2003, Griffith2004}, as discussed in section \ref{s:models}.

Similar to CO$_2$, the Cassini flyby provided more complete latitudinal coverage of the HCN column density than that provided by the observations of \citet{Griffith2004}, and at higher spatial resolution than observations of \citet{Moreno2003}. The Cassini-CIRS observations measured the HCN column density as a function of latitude 6.5\,years post SL9 \citep{Lellouch2006}. Not only did it allow estimation of the strength of the vertical and meridional mixing, but it also revealed a sharp drop by a factor $\sim$7 from the SL9 impact latitude to the south polar region. This drop was hard to reconcile with the CO$_2$ polar excess, given the simultaneity of these observations, combined with the fact that both species were presumed to derive from the SL9 impact. The conclusion was that the HCN distribution could be explained assuming a sharp cutoff in the meridional eddy mixing coefficient (over an order of magnitude) poleward of 40$^{\circ}$S combined with equatorward advective transport. The markedly different distribution of CO$_2$ and HCN led \citet{Lellouch2006} to conclude that they likely resided at different pressure levels (5-10\,mbar and 0.5\,mbar, respectively), and therefore implying very different horizontal transport regimes.

\begin{figure}[!h]
 \begin{center}
    \includegraphics[width=0.9\textwidth]{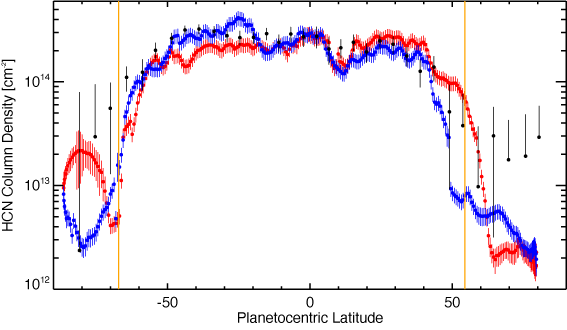}
  \end{center}
  \caption{HCN column density derived from ALMA at the eastern (red) and western (blue) limbs observed on in March 2017 (22.5\,years post-SL9) \citep{Cavalie2023b}. It is compared with the CIRS-derived HCN column density in December 2000 (6.5\,years post-SL9) shown in black \citep{Lellouch2006} and divided by a factor of 10. The vertical orange lines represent the boundary of the auroral regions.}
  \label{fig:HCN}
\end{figure}

ALMA observations recorded in March 2017 provided another measure of the HCN meridional distribution 22.5\,years after the SL9 impact \citep{Cavalie2023b}, shown here on Fig.~\ref{fig:HCN}. Thanks to ALMA's high-spectral resolution, one notable feature from this dataset was the strikingly different lineshape of the HCN (4-3) emission line between the high- and low-latitude regions. This indicates that the vertical distribution of HCN is different between these two regions, with HCN residing at p $<$ 0.1\,mbar in the auroral regions, and p $<$ 3\,mbar at lower latitudes. 

Two additional striking features were also discovered in the ALMA dataset from \citet{Cavalie2023b}. First is the factor of 10 higher HCN column density observed at 75$^{\circ}$S–85$^{\circ}$S, around 350$^{\circ}$W compared to the other measurements at nearby latitudes, and interpreted as a high-altitude HCN production source within the auroral regions. Second is the almost 2 orders of magnitude drop in the HCN column density above both poles compared to the low latitudes, that they interpreted as a result from heterogeneous chemistry with stratospheric aerosols, and which will be discussed in section \ref{ss:aerosols}.

\subsection{Ions}
\label{ss:Ions}

The only ion detected in Jupiter's atmosphere is H$_3^+$, first discovered by \citet{Drossart1989}. Charged particle precipitations in the upper atmosphere of Jupiter lead to the formation of H$_2^+$ through electron impact ionization H$_2$ + e$^{-}$ $ \rightarrow$ H$_2^+$ + 2\,e$^{-}$. This light ion then reacts with molecular hydrogen to form H$_3^+$ through the reaction:

\begin{equation}
     \mathrm{H}_2^+ + \mathrm{H}_2 \rightarrow  \mathrm{H}_3^+ + \mathrm{H} \label{eq:h3+}
\end{equation}

H$_3^+$ produces emission lines in the near-IR, which allows the ionosphere of Jupiter to be sounded. The emission lines can be inverted using radiative transfer calculations to derive H$_3^+$ temperature and number density, assuming the ion is in quasi-local thermodynamical equilibrium (LTE) \citep[see, e.g., the review of][]{Badman2015}. This ion number density peaks around the auroral regions and decreases towards lower latitudes along with this species' temperature, showing it plays an important role in controlling the upper atmospheric temperature of Jupiter \citep{ODonoghue2021}.

H$_3^+$ is lost through dissociative recombination above the methane homopause. The lifetime of the H$_3^+$ electron recombination is given by $\tau_r$ = ($k_r$ $n_{e}$)$^{-1}$, where $k_r$ is the recombination rate constant and $n_{e}$ the electron density \citep{Achilleos1998}. In the auroral region, the electron densities derived from the Voyager 2 radio occultation measurement at 66$^{\circ}$S, 258$^{\circ}$W range from $\sim$5$\times$10$^{4}$ to 3$\times$10$^{5}$\,cm$^{-3}$ in the 600-1200\,km altitude range \citep{Hinson1998}, i.e., where the H$_3^+$ emission peak occurs \citep{Tao2011}. Using the formula for $k_r$ =  1.15\,$\times$\,10$^{-7}$\,(300/T)$^{0.65}$\,cm$^{3}$s$^{-1}$ from \citet{Sundstrom1994}, this leads to an H$_3^+$ lifetime in the 20-300\,s range. This lifetime is orders of magnitude shorter than the lifetime of the neutral chemical species discussed in this paper, and therefore measurements of the H$_3^+$ distribution can only be used to trace dynamical processes over a relatively short timescale, as discussed in section \ref{s:coupling}.

Note that, below the methane homopause, H$_3^+$ is also destroyed through reaction with CH$_4$ (H$_3^+$ + CH$_4$ $\rightarrow$ CH$_4^+$ + H$_2$) and heavier hydrocarbons, such as C$_2$H$_2$, C$_2$H$_4$, and C$_2$H$_6$ \citep{Kim1994}, as detailed in section \ref{ss:ion_neutral_models}.

\subsection{Aerosols and hazes}
\label{ss:aerosols}

The term hazes has been historically used to define layers of sub micron-sized particles located in the upper troposphere (200-500\,mbar), as well as stratospheric particles located at pressures lower than 100\,mbar, while the term cloud refers to larger-sized condensates composing the more variable layers seen in the visible at higher pressure levels (see reviews of \citealt{West1986, West2004}). 

\begin{figure}[!h]
 \begin{center}
    \includegraphics[width=0.95\textwidth]{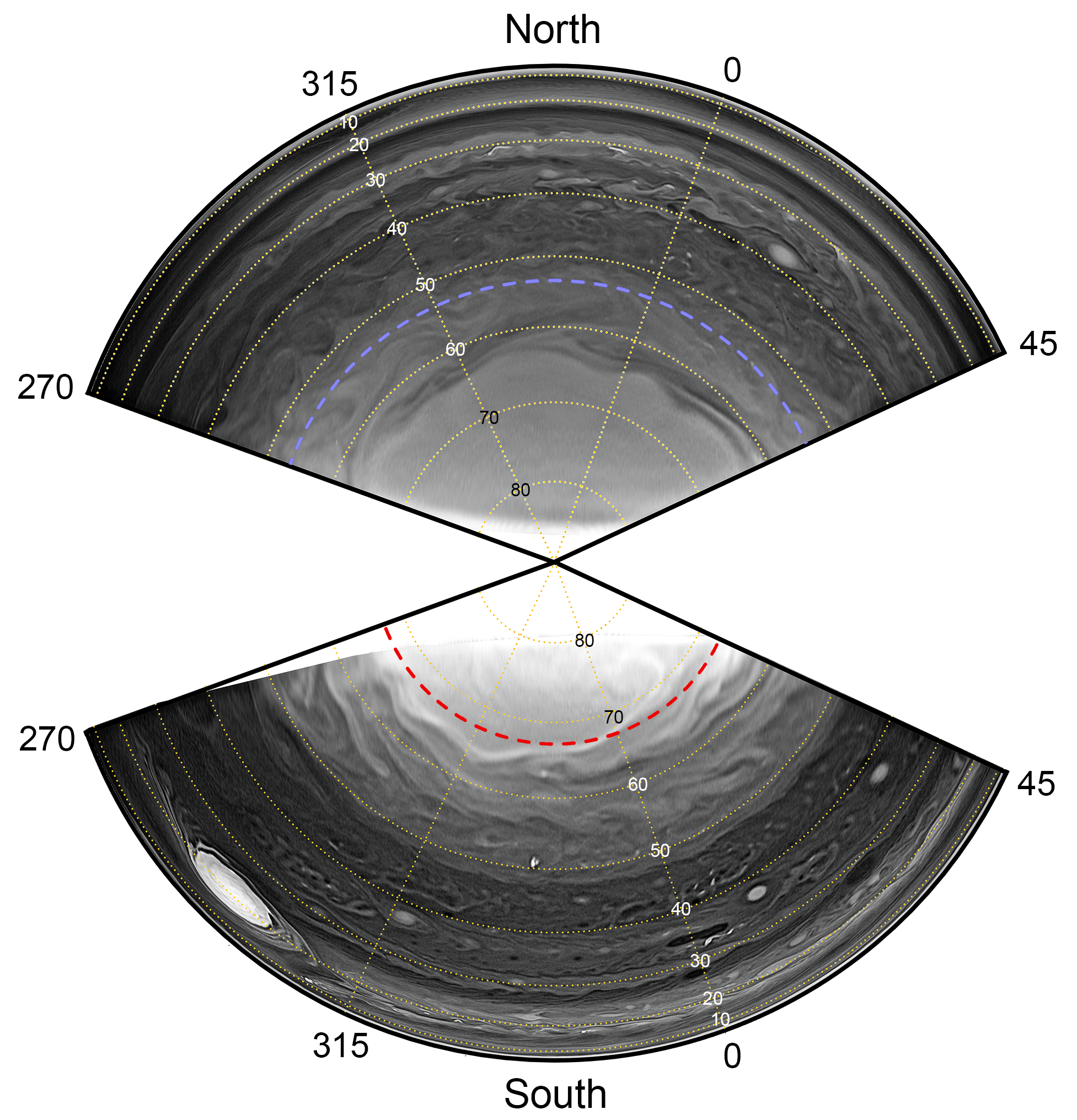}
  \end{center}
  \caption{Composite image of James Webb Space Telescope observations of Jupiter in a narrow 2.12\,$\mu$m filter (F212N; data from \citealt{Hueso2023}). The image has been contrast-enhanced and high-pass filtered to show the complex details present in the hazes. Following Fig. \ref{fig:Giles2023_vs_Cavalie2023}, red/blue dashed lines in the southern/northern hemisphere show the lowest parallel reached by the expanded auroral reference ovals.}
  \label{fig:polar_methane_band}
\end{figure}

Some of the most striking high-latitude features at Jupiter's poles include a near-UV dark polar cap, and the bright polar hood as seen in near-IR methane band imagery \citep[e.g.][and references therein]{West2004}, from which the aerosol distribution, size and shape can be constrained using observations at multiple phase angles. Figure \ref{fig:polar_methane_band} shows snapshots of the bright polar hoods in a strong methane absorption band at 2.1\,$\mu$m, as captured by JWST/NIRCam \citep{Hueso2023}. The hazes are more optically thick in the south hemisphere than in the north. In both hemispheres, the latitudinal distributions of the lower stratospheric hazes are constrained by defined boundaries showing a complex wave structure consistent with Rossby waves \citep{SanchezLavega1998, BarradoIzagirre2008}, and JunoCam provided some of the most spatially-resolved observations of this wave pattern to date \citep{Rogers2022}.

Cloud structure and formation, as well as upper tropospheric haze layers will not be discussed here, as the focus of this paper is on the distribution of stratospheric species with an emphasis on the polar regions. However, we refer the reader to the work of \citet{Wong2020} that serves as a reference to the observing campaigns performed simultaneously to the Juno mission from 2016-2019, in the UV/Visible/near-IR, providing information on the upper-tropospheric wind field, as well as cloud structure and distribution.

\citet{Pryor1991} used observations from the Voyager 2 Photopolarimeter Subsystem (PPS) to characterize and provide an explanation for Jupiter's UV-dark polar cap, previously discovered by \citet{Hord1979}. They compared the location where this behavior was observed with Jupiter's auroral zone, and suggested that ion-neutral chemistry would lead to the destruction of methane, and ultimately promote the creation of haze through the production of large hydrocarbons.

\citet{Kim1991} used spectroscopic observations in the broad 2-$\mu$m methane absorption band to constrain the altitudes of the polar hazes to pressure levels from 70 to 5\,mbar. \citet{SanchezLavega1998} used HST and Voyager observations in the UV and methane absorption band at 890\,nm to characterize the polar stratospheric hazes, showing that the hazes are located at around 100\,mbar and are latitudinally limited by Rossby waves that interact with the zonal jets at the troposphere. \citet{Vincent2000} used HST observations in the UV to study the polar stratospheric aerosols at pressures of few tens of mbar, finding evidence of weak westward motions in the range of -3 to -11 m\,s$^{-1}$ and meridional mixing in the polar stratosphere.

\citet{Rages1999} used Galileo observations recorded by its Solid State Imaging subsystem (SSI) in the violet ($\sim$417\,nm) and near-IR ($\sim$756\,nm) to study the stratospheric hazes. The SSI violet observations are sensitive to the particle properties around the 20\,mbar pressure region and above. They retrieved stratospheric aerosol particles with sub-micron sizes, and haze number densities increasing by one order of magnitude from low- to high-latitudes. They also obtained an alternative solution at high latitude, with particle sizes in the 1.3\,$\mu$m range.

The Cassini/ISS instrument provided images of Jupiter through several filters at a wide range of phase angles. Some of the filters of interest for this present paper are the UV1 (0.258\,$\mu$m), MT3 (0.889\,$\mu$m) and CB3 (0.938\,$\mu$m) filters. The UV1 filter probes the lower stratosphere due to Rayleigh scattering being the main opacity source. Methane absorption is the main opacity source in the wavelength region covered by the MT3 filter which probes the lower stratosphere, while CB3 is a continuum filter centered around the methane window \citep{Porco2003, Porco2004}. These observations resulted in detailed maps of the polar aerosols \citep{BarradoIzagirre2008}, and showed the formation and later evolution and dispersal of a large dark oval inside the north auroral oval probably associated with an auroral event \citep{Barbara2024}.

Several additional studies have attempted to produce detailed retrievals of the properties of the polar stratospheric aerosols. Post-SL9 observations in the near-IR H (1.45-1.8$\mu$m) and K (1.95-2.5$\mu$m) bands were previously used to retrieve the aerosol properties and distribution on Jupiter \citep{Banfield1996}. Using this inversion method, \citet{Banfield1998b} extended the retrieval to a larger range of latitudes. They calculated the location of the haze layer at low-latitudes to be $\sim$50\,mbar while, in the polar region, their location was estimated to be $\sim$20\,mbar. Some of the assumptions made in these works include holding the aerosol particle size constant (0.3\,$\mu$m), i.e., within the particle size range of 0.2-0.5$\mu$m, as previously constrained using near-UV observations \citep{Tomasko1986}.

While the 0.3\,$\mu$m-size aerosols provide a good fit of the near-IR at low-latitudes, reproduction of the 1.7-1.75\,$\mu$m and 2.05-2.1\,$\mu$m ranges at high latitude proved more challenging \citep{Banfield1998b}. Using a similar method to \citet{Banfield1996, Banfield1998b}, \citet{Zhang2013b} relaxed the constraints on the aerosol particle size and demonstrated that the high-latitude spectra were better fit when considering 0.7\,$\mu$m-radius aerosols. \citet{Zhang2013b} also improved the method from \cite{Banfield1996} by using refined CH$_4$ absorption coefficients from \citet{Karkoschka2010}, which slightly shifts to higher pressure levels the aerosol layer compared to the work of \citet{Banfield1998b}. The inferred stratospheric aerosol map of \citet{Zhang2013b} displayed a dichotomy between mid- and high-latitude. At low- to mid-latitudes, stratospheric aerosols are located within 1-2 scale heights near 50\,mbar, while they are above 20\,mbar at higher latitudes. 

\begin{figure}[!h]
 \begin{center}
    \includegraphics[width=0.99\textwidth]{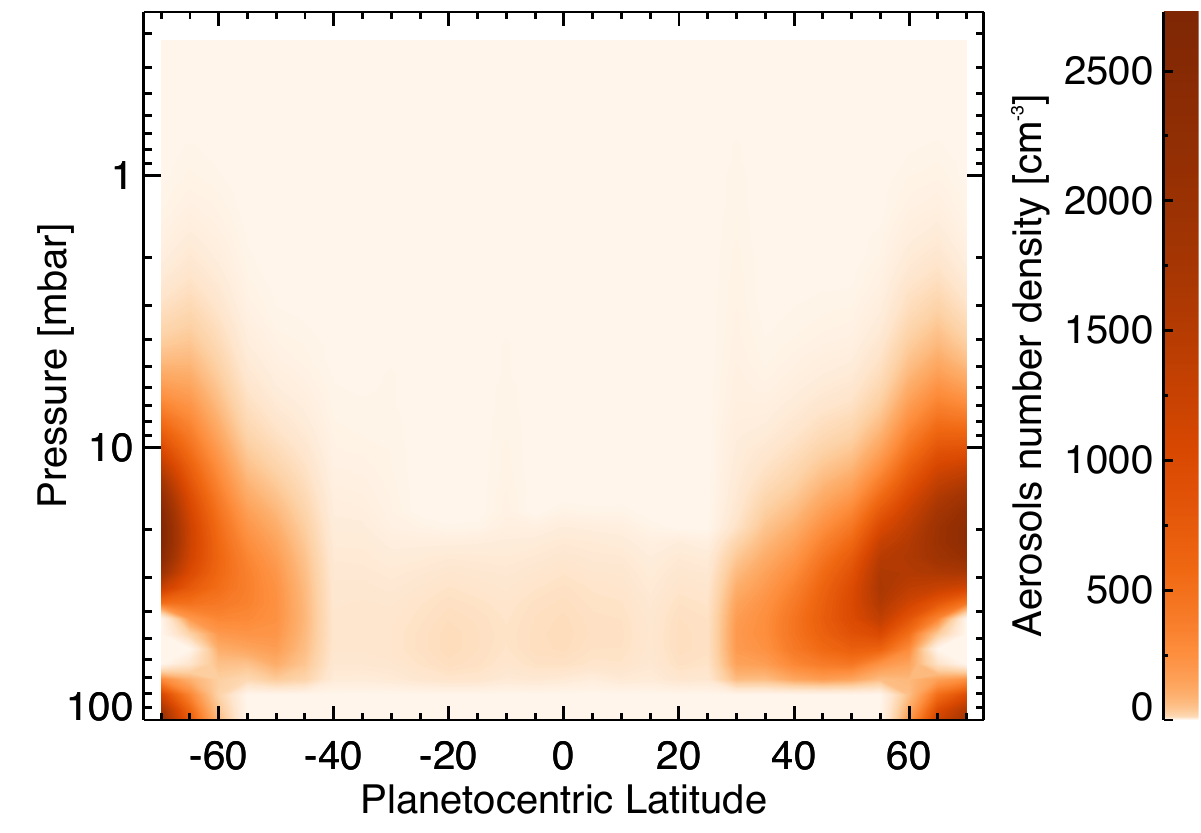}
  \end{center}
  \caption{Zonally-averaged aerosol map (cm$^{-3}$) derived from ground-based near-IR observations and Cassini/ISS observations from \citet{Zhang2013b}.}
  \label{fig:aerosol_map}
\end{figure}

By combining ground-based near-IR observations and the Cassini/ISS limb-darkening observations at multiple phase angles, \citet{Zhang2013b} were able to derive an aerosol density map. They tested two properties of aerosols; one consisting of compact sub-micron particles, and the other consisting of fractal aggregates composed of small monomers. While the scattering properties of compact sub-micron particles are best reproduced using Mie theory, additional phase functions (e.g., Henyey-Greenstein) are needed for larger particles. They conclude that two types of particles are needed to explain the UV1 and MT3 limb-darkening profiles recorded at multiple phase angles. The low-latitude (40$^{\circ}$S-25$^{\circ}$N) images are best fit using compact sub-micron particles, with sizes in the 0.2-0.5$\mu$m range. The high-latitude regions are best fit assuming fractal aggregate particles consisting of thousands of $\sim$10\,nm monomers. The derived aerosol column density is $\sim$2 orders of magnitude greater in the polar versus the low-latitude regions. While they noted the non-uniqueness of their solution, they provided the possible range of monomer size and number depending on the assumed imaginary part of the UV refractive index. Fig. \ref{fig:aerosol_map} shows the aerosol distribution map retrieved by \citet{Zhang2013b}.

\begin{figure}[!h]
 \begin{center}
    \includegraphics[width=0.9\textwidth]{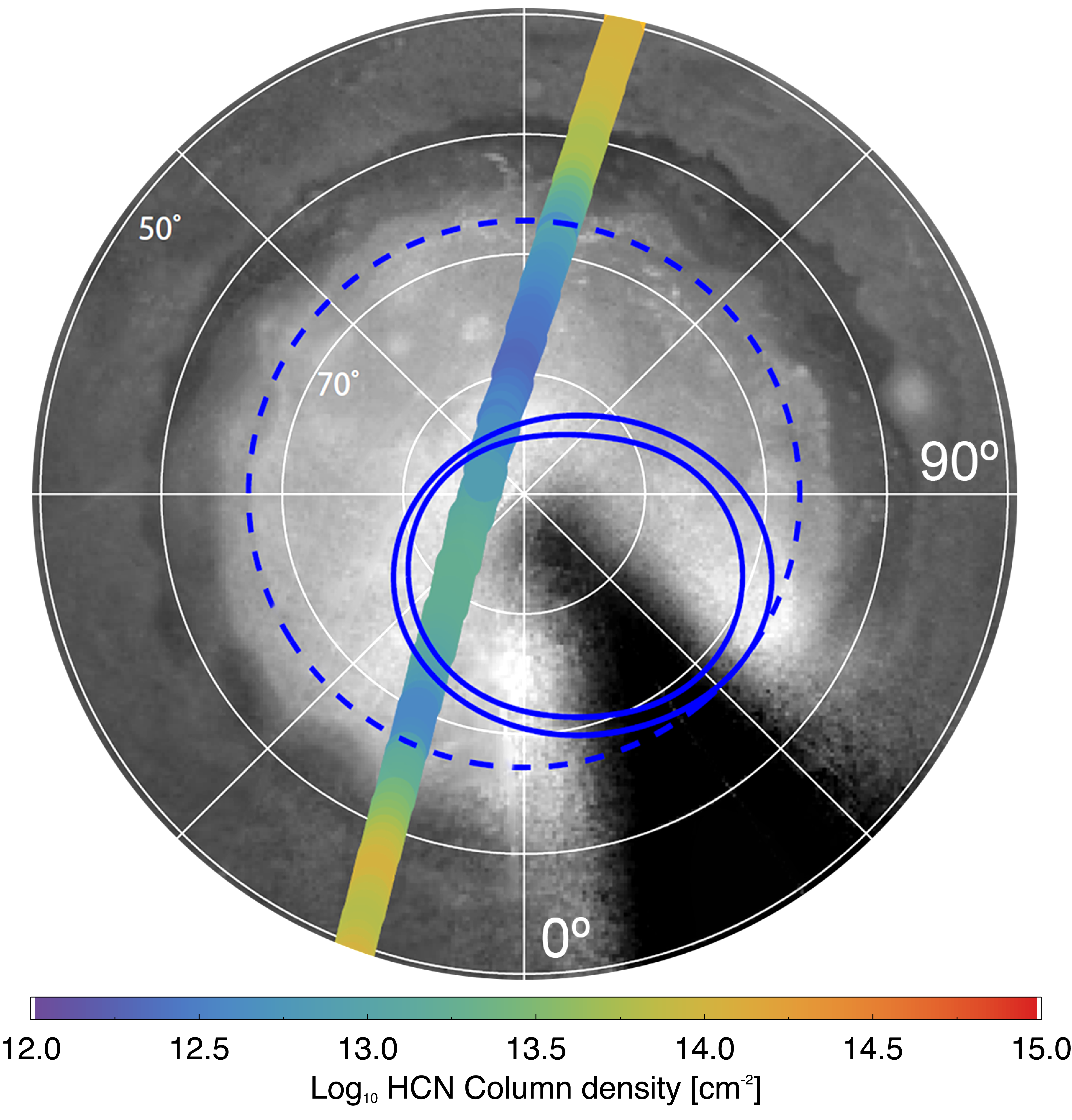}
  \end{center}
  \caption{Composite image of the methane-filter image of Jupiter's south polar region from JunoCam \citep{Rogers2022} recorded during Juno PJ24, and compared with the HCN column density using ALMA derived by \citet{Cavalie2023b}. The contracted and expanded reference auroral ovals derived from HST observations are displayed in solid blue line \citep{Bonfond2012a}. The lowest parallel reached by the expanded southern reference oval (67$^{\circ}$S) is displayed as a dashed circle.}
  \label{fig:HCN_vs_aerosol}
\end{figure}

Heterogeneous chemistry on aerosols may affect the chemical distribution of several species. Laboratory experiments found that HCN bonds with aerosols once they reach a typical size of $\sim$0.1\,$\mu$m \citep{Perrin2021}. Such sized aerosols are predicted to be present in Jupiter's polar regions around the $\sim$mbar level from observations \citep{Zhang2013b}, and modeling work \citep{Friedson2002}, as will be described in section \ref{s:models}. Figure \ref{fig:HCN_vs_aerosol} compares the methane-band image of the JunoCam experiment \citep{Hansen2017}, with the HCN column density derived by \citet{Cavalie2023b}. In the absence of clouds, JunoCam methane-filter images probe a pressure level of 540\,mbar, as determined from the $\tau$ = 1 optical depth at 889\,nm from \citet{SanchezLavega2013}, and higher in the stratosphere in the presence of clouds, and may be used as a proxy for the aerosol distribution. The striking spatial correlations between the HCN depletion in the region of enhanced hydrocarbon production near the auroral region (Fig. \ref{fig:Giles2023_vs_Cavalie2023}) and polar aerosols (Fig. \ref{fig:HCN_vs_aerosol}) favor the scenario where HCN is depleted from heterogeneous chemistry, although more experimental work is needed to constrain the heterogeneous chemical reactions involved.

Reviews of several relevant models aiming at explaining the observed chemical distributions are now summarized.

\section{Atmospheric models}
\label{s:models}

\subsection{Neutral photochemical models}
\label{ss:neutral_models}

Methane photolysis and its consequence on stratospheric chemistry caught the interest of the community several decades ago \citep{Cadle1962, Strobel1969, Strobel1973}. In the mid-IR, hydrocarbons are detected in the mbar and sub-mbar levels. In these regions, their abundances are controlled by a combination of chemical reactions triggered by the by-products of methane photolysis and subsequent chemistry, combined with downward diffusion as well as horizontal mixing.

\citet{Allen1981} initially developed the Caltech/JPL 1D photochemical and transport Model (KINETICS), which was further refined later \citep{Yung1984, Gladstone1996, Moses2000a, Moses2005a}. Alternatively, \citet{Dobrijevic1998} developed a photochemical model for the giant planets, not only to interpret ground- and space-based observations, but also to study the sensitivity of the model output to uncertainties in the chemical reaction rates \citep{Dobrijevic2003, Dobrijevic2010, Dobrijevic2011}. 

Photochemical and thermochemical models allow the in-depth study of the chemical inventory in planetary atmospheres. Unlike general circulation models, photo-/thermo-chemical models are restricted to solving the continuity equation for a set of non-linearly coupled equations, each equation representing the temporal evolution of a given individual species. This allows exploring the complexity of the chemical network, despite having simplified transport processes. Because several hydrocarbons were detected early on, these models initially focused on reproducing the hydrocarbon distributions. 

One-dimensional (altitude) photochemical models were first developed to interpret the early observations of the outer planets, since such observations were mostly disk-averaged, or with a handful of resolution elements. Such models would generally solve $i$ continuity equations (\ref{eq:PM_1D}), non-linearly coupled by a set of $m$ reactions:

\begin{equation}
    \dfrac{ \partial n_i}{\partial t} = P_i - n_i\,L_i - \dfrac{\partial \Phi_i}{\partial z} \label{eq:PM_1D}
\end{equation}

\noindent 
where $n_i$, $z$, $P_i$, $L_i$, and $\Phi_i$ represent the abundance of species $i$, the altitude, the chemical production term of species $i$, the chemical loss term of species $i$, and the vertical flux of species $i$, respectively. Depending on the number of species considered and the extent of the network, $i$ generally ranges from a few dozen to several hundred, while $m$ can be up to several thousand.

The observations gathered by missions such as Cassini-Huygens, or the constantly increasing ground-based capabilities provided the abundance distribution of certain species as a function of latitude, longitude and season, therefore challenging the predictive capabilities of uncoupled 1D-models covering different latitudes. Using Saturn as an illustration, \citet{Moses2005b} first developed a seasonal model of Saturn's stratosphere, consisting of a series of 1D-models, in order to interpret the IRTF/TEXES mid-IR observations of Saturn \citep{Greathouse2005}. They showed that the meridional trend of C$_2$H$_2$ was fairly well reproduced by the model, characterized by a meridional decrease in its abundance from equator to higher latitudes, thus following the yearly average of solar illumination. However, their predicted C$_2$H$_6$ meridional trend was anti-correlated with the observations, likely due to the higher sensitivity of C$_2$H$_6$ to transport processes because of its longer chemical lifetime. 


Cassini-Huygens arrived at Saturn in July 2004, and the CIRS instrument provided a comprehensive dataset of nadir- and limb-sounding measurements of Saturn's main hydrocarbons at different latitudes and over half a Kronian year \citep{Flasar2005, Howett2007, Guerlet2009, Guerlet2010, Sinclair2013, Sylvestre2015}. To interpret these observations, \cite{Hue2015, Hue2016} developed a seasonal 2D-photochemical (altitude-latitude) model that calculates the seasonal variation of abundances, first by turning the meridional diffusion off. Essentially, the set of equations solved in a 2D-photochemical model (see Eq. \ref{eq:PM_2D}) is similar to the 1D model:


\begin{equation}
    \dfrac{\partial n_i}{\partial t} = P_i - n_i\,L_i - \dfrac{1}{r^2}\dfrac{\partial ( r^2 \Phi_i^r )}{\partial r} + \dfrac{1}{r \cos{\theta}} \, \dfrac{ \partial ( \cos{\theta} \Phi_i^{\theta}) }{\partial \theta}\label{eq:PM_2D}
\end{equation}

\noindent 
where $r$, $\theta$, $\Phi_i^{r}$, $\Phi_i^{\theta}$ represent the radial direction, the latitude, the vertical flux of species $i$, and the meridional flux of species $\theta$, respectively. 

Cassini/CIRS limb-sounding measurements on Saturn were reasonably well-reproduced from the equator up to $\pm$40$^{\circ}$ in the 0.1-1\,mbar range. Deviation from the model prediction beyond 40$^{\circ}$S indicates large-scale stratospheric dynamics \citep{Hue2015}. Some of the main takeaway from the \citet{Moses2005b} and \citet{Hue2015} studies is that photochemistry controls the abundance distribution of the C$_2$H$_\mathrm{x}$, and that the deeper in the stratosphere the more they are controlled by the annually-averaged insolation instead of the instantaneous insolation. The predicted seasonal variability in the C$_2$H$_\mathrm{x}$ molecules are only noticeable at pressures lower than 0.1\,mbar.

On Jupiter, the situation is different due to its lower obliquity, making seasonal effects more muted than on Saturn. \citet{Liang2005} developed a quasi-two-dimensional model to study the effect of meridional diffusion on the hydrocarbon meridional distributions from Cassini/CIRS \citep{Kunde2004}. They coupled the 1D results from KINETICS with a meridional diffusion coefficient, K$_{yy}$, and inferred a K$_{yy}$ $\sim$ 10$^{11}$\,cm$^{2}$\,s$^{-1}$ above the 10\,mbar level, and K$_{yy}$ $\leq$ 10$^{9}$\,cm$^{2}$\,s$^{-1}$ below the 5-10\,mbar pressure level, in agreement with post-SL9 dust tracking, though about two orders of magnitude smaller than values derived from tracking post-SL9 stratospheric chemical species \citep{Moreno2003, Griffith2004, Lellouch2006}.

Diffusion models have been developed to interpret the post-SL9 species, bringing constraints on diffusive processes occurring in Jupiter's stratosphere. Horizontal diffusion models have been developed to understand the latitudinal spread of the post-SL9 species such as HCN, CO, CS \citep{Moreno2003, Griffith2004}, providing an estimate of the K$_{yy}$ magnitude. Such approaches assume that the vertical diffusion timescale is longer than the horizontal one, and that these species are chemically stable, which are both reasonable assumptions \citep{Moses1995}. 

\citet{Lellouch2006} expanded upon this approach and modelled the distribution of HCN and CO$_2$ (see Figs. \ref{fig:Oxy} and \ref{fig:HCN}) including a meridionally variable K$_{yy}$ and meridional winds v$_{\theta}$. They used an inversion technique to derive the optimal parameter K$_{yy}$($\theta$) first by only assuming meridional diffusion, and then by adding advective transport. Considering HCN first, they added a HCN chemical production term, deriving from the NH$_3$ photolysis. They concluded that (i) HCN production from NH$_3$ is negligible in controlling its meridional distribution, (ii) equatorward winds are needed to reproduce the HCN observations. They proceeded with modeling the CO$_2$ distribution, while accounting for the relevant chemical species and reactions involved in the production and destruction of CO$_2$, based on \citet{Lellouch2002}. They were successful at reproducing the CO$_2$ polar enhancement assuming meridionally uniform K$_{yy}$ combined with a poleward advective transport. The difference in the advective transport needed to reproduce HCN (equatorward transport) and CO$_2$ (poleward transport) was interpreted as an indication that both species resided at different pressure levels, i.e., 5-10\,mbar for CO$_2$ and 0.5\,mbar for HCN (as discussed in section \ref{ss:Cyanide}).

\citet{Hue2018} adapted a version of their Saturn seasonal model to Jupiter in order to compare their results with the hydrocarbon measurements retrieved from the Cassini/CIRS observations \citep{Nixon2010}. The use of a 2D model allows the addition of K$_{yy}$ as well as 2D wind patterns. The meridional distributions of C$_2$H$_2$ and C$_2$H$_6$ measured by Cassini/CIRS were compared to the model output, while considering the following number of cases: \\
\indent
$\bullet$ K$_{yy}$ = 0 (equivalent to a sum of 1D models run at different latitudes), \\
\indent
$\bullet$ K$_{yy}$ $\neq$ 0 (using constraints from post-SL9 chemical species tracking of \citet{Lellouch2002, Moreno2003, Griffith2004, Lellouch2006}), \\
\indent
$\bullet$ K$_{yy}$ $\neq$ 0 and advective circulation (winds).\\

\begin{figure}[!h]
 \begin{center}
    \includegraphics[width=0.7\textwidth]{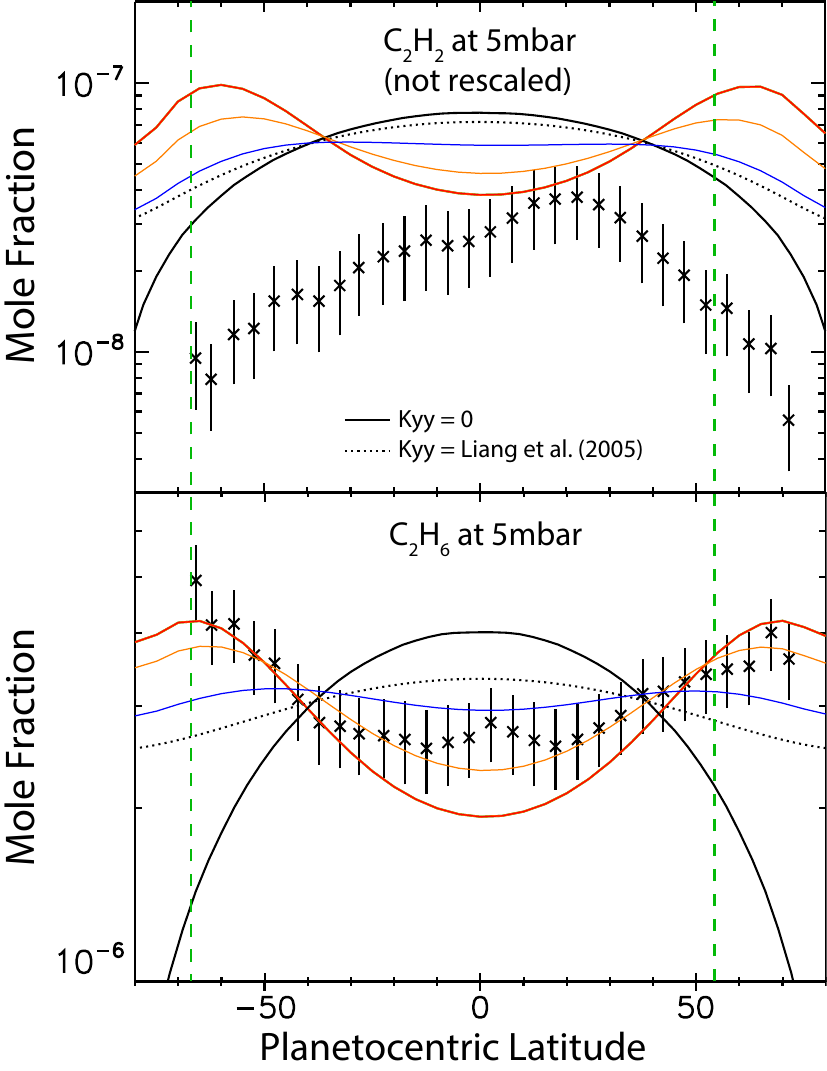}
  \end{center}
  \caption{Meridional distribution of C$_2$H$_2$ and C$_2$H$_6$ abundances at 5\,mbar from \citet{Hue2018}. Black solid line: photochemical predictions with K$_{yy}$ = 0. Black dotted line: photochemical predictions with K$_{yy}$ taken from \cite{Liang2005}. Red/blue/orange lines: photochemical predictions with K$_{yy}$ taken from \cite{Liang2005} and stratospheric 2D advective transport. The model results are compared with the Cassini/CIRS observations of Jupiter \citep{Nixon2010}. The lowest parallels reached by the expanded reference ovals are displayed as vertical dashed green lines. Figure adapted from \citet{Hue2018}.}
  \label{fig:C2H2_C2H6_hue}
\end{figure}

Results of a 2D-photochemical model that assumes a constant K$_{yy}$ with altitude (K$_{yy}$ = 2\,$\times$\,10$^{11}$\,cm$^{2}$s$^{-1}$), combined with a two-cell circulation pattern are shown in Figure \ref{fig:C2H2_C2H6_hue}, taken from \citet{Hue2018}. The circulation cells were parameterized assuming upwelling winds at low-latitudes ($<$30$^{\circ}$) and downwelling winds at higher latitudes. The two nodes of the cells were arbitrarily fixed at $\pm$30$^{\circ}$, as a numerical experiment, in order to demonstrate the interplay between the strength of the advective (winds) and diffusive (K$_{yy}$) transport in controlling the meridional distribution of hydrocarbons. \citet{Hue2018} used constraints from \cite{Lellouch2006} to build their 2D-wind field. \citet{Lellouch2006} assumed a sinusoidal variation in the amplitude of the latitudinal winds, with maximum winds at $\pm$45$^{\circ}$ latitude and zero at the equator and the poles (see their Figure 7). In the southern hemisphere at $\pm$45$^{\circ}$ latitude, they assumed winds of 7-9\,cm\,s$^{-1}$ at $\sim$\,0.5\,mbar towards the equator, and winds of 30\,cm\,s$^{-1}$ in the direction of the poles at $\sim$\,5-10\,mbar to reproduce the HCN and CO$_2$ distribution, respectively.

\begin{figure}[!h]
 \begin{center}
    \includegraphics[width=0.98\textwidth]{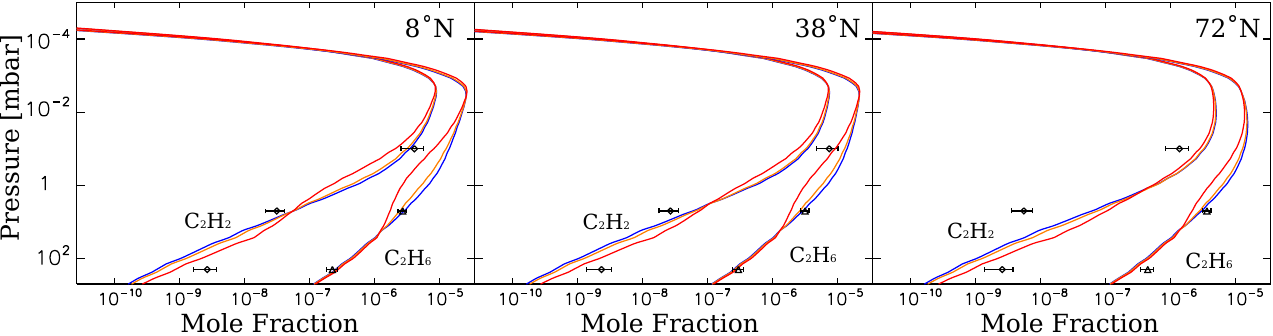}
  \end{center}
  \caption{Vertical profiles of C$_2$H$_2$ and C$_2$H$_6$ at 8N (left panel), 38N (middle panel) and 72N (right panel). The vertical profiles are only presented when the stratospheric circulation is added to the photochemical model. The corresponding wind velocities are presented on Fig. 9 of \cite{Hue2018}. The model results are compared to the Cassini/CIRS observations of Jupiter at 0.1 mbar (C$_2$H$_2$ only), 5 mbar and 200 mbar \cite{Nixon2010}. Image reproduced with permission from \citet{Hue2018}, copyright by Elsevier.}
  \label{fig:C2H2_C2H6_vertical_hue}
\end{figure}

\citet{Hue2018} showed that, in order to reproduce the 5\,mbar C$_2$H$_6$ meridional trend measured by Cassini/CIRS, a circulation cell flowing in the opposite direction, i.e., with upwelling motions at the equator and downwelling motion at higher latitude, was needed. Model outputs are displayed on Figures \ref{fig:C2H2_C2H6_hue} and \ref{fig:C2H2_C2H6_vertical_hue}, and compared with the Cassini/CIRS measurements from \citet{Nixon2010}. Figure \ref{fig:C2H2_C2H6_hue} shows that while the C$_2$H$_2$ meridional gradient can be reproduced using K$_{yy}$ = 0 and no winds (solid line), the meridional distribution of C$_2$H$_6$ is however best reproduced, in terms of equator-to-pole gradients, using a combination of moderate meridional diffusion coefficient (K$_{yy}$ $\neq$ 0) with a circulation cell with upwelling motions at the equator and downwelling motions at mid-latitudes. Alternatively, implementing in the 2D-photochemical model a circulation cell using loose constraints suggested by \citet{Lellouch2006} reinforces the equator-to-pole C$_2$H$_2$ meridional gradient, but degrades at the same time the model/observation agreement for C$_2$H$_6$. 





Compared to Earth's atmospheric circulation, the understanding of transport mechanisms in Jupiter's stratosphere is limited. Voyager/IRIS performed the first spatially resolved measurements of the latitudinal distributions of wind and temperature. These measurements allowed \citet{Gierasch1986} to deduce the existence of a circulation in the upper troposphere, using an axisymmetric model that accounts for perturbations in the zonal wind and thermal fields. \citet{Conrath1990} extended this study by including radiative forcing and extending the calculation to the stratosphere. In the stratosphere, the predicted meridional circulation turns into two large cells with upwelling at the equator and downwelling at the poles. However, their calculations did not account for atmospheric aerosols. 

\citet{West1992} and \citet{Moreno1997} demonstrated that including aerosols, with different distributions, may lead to opposite circulation patterns in the stratosphere, suggesting the atmospheric circulation is highly sensitive to the aerosol properties. \citet{Guerlet2020} and \citet{Zube2021} highlighted the critical role of assumed aerosol distributions in controlling the Jovian stratospheric radiative budget and circulation patterns. \citet{Guerlet2020} found that, at pressure less than 3 mbar, the predicted circulation is independent of aerosol assumptions, with upwelling motions near the equator and downwelling poleward of 50$^{\circ}$. In the lower stratosphere, the predicted circulation of \citet{Guerlet2020} varies with the considered aerosol scenarios, and the predicted circulation with their most realistic scenario would be generally inconsistent with the \textit{ad hoc} circulation implemented by \citet{Hue2018} to reproduce the Cassini/CIRS observations.

Because C$_2$H$_2$ and C$_2$H$_6$ are chemically coupled and have lifetimes of the same order of magnitude, \citet{Hue2018} showed that no combination of diffusion coefficients ($K_{zz}$ and $K_{yy}$) with a circulation pattern can explain the meridional distributions CIRS recorded. A alternative way to produce radically different meridional distributions for these two compounds would be to invoke a (photo-)chemical process that would be strongly coupled to only one of them. Ion-neutral chemistry could be a good candidate, and will be discussed in the next section. However, one should note that the retrieval performed on the Cassini/CIRS observations by \citet{Nixon2007, Nixon2010} excluded the longitudinal range around the auroral region.

\subsection{Ion-Neutral photochemical models, and aerosol formation models}
\label{ss:ion_neutral_models}

\cite{Kim1994} initially calculated the effect of photo-ionization on the hydrocarbon ion chemistry. The ionization produces secondary electrons which, in turn, can ionize, dissociate and excite atmospheric gases. They developed an atmospheric model separated into two regions, (i) the ionosphere above 700\,km (2$\times$10$^{-6}$\,mbar), where only light ions containing H and He atoms were included, as well as atomic hydrogen, and (ii) the lower ionosphere below 700\,km, where hydrocarbons up to two carbons were included, and pseudoions for the C$_3$H$_\mathrm{x}$ and C$_4$H$_\mathrm{x}$. Note that \cite{Kim1994} are using the 0.6\,bar ammonia cloud top as the z\,=\,0\,km reference altitude. We have added the corresponding atmospheric pressure associated with each result from their paper. The lower ionospheric model solves the continuity equation for 21 ions and 12 neutrals. The abundances of CH$_4$, C$_2$H$_2$, and C$_2$H$_6$ at the lower boundary of the lower ionospheric model were set consistently according to Voyager UV occultation data, while the abundances of CH$_3$ and C$_2$H$_4$ were set according to photochemical model calculations from \citet{Gladstone1996}. The two models were solved iteratively and the results at the common boundary (700\,km) were fed into each other until convergence was reached.


\citet{Kim1994} predicted the formation of a hydrocarbon ion layer around 270-470\,km altitude ($\sim$ 10$^{-2}$-10$^{-4}$\,mbar). CH$_5^{+}$, C$_2$H$_3^{+}$, CH$_3^{+}$ and C$_2$H$_5^{+}$ are the dominant light hydrocarbon ions in the altitude range of 360-430\,km, corresponding to pressures of $\sim$ 10$^{-3}$-2$\times$10$^{-4}$\,mbar. CH$_5^{+}$ is destroyed by dissociative recombination while CH$_3^{+}$ reacts with H$_2$ to produce CH$_5^{+}$ or undergoes dissociative recombination, at 400\,km (7$\times$10$^{-4}$\,mbar). C$_2$H$_\mathrm{x}$ ions are produced at 350-420\,km altitude (3$\times$10$^{-3}$-4$\times$10$^{-4}$\,mbar), where reactions with methane (CH$_3^{+}$ + CH$_4$ $\rightarrow$ C$_2$H$_5^{+}$ + H$_2$) and acetylene (CH$_5^{+}$ + C$_2$H$_2$ $\rightarrow$ C$_2$H$_3^{+}$ + CH$_4$) were found to be the dominant ones above 350\,km (3$\times$10$^{-3}$\,mbar). Below 350\,km, the dominant pathway to C$_2$H$_\mathrm{x}$ ions production becomes photoionization of C$_2$H$_2$, C$_2$H$_4$, and C$_2$H$_6$. Near 320\,km (7$\times$10$^{-3}$\,mbar), C$_2$H$_\mathrm{x}$ ions are found to be efficiently converted into C$_3$H$_\mathrm{x}$ and C$_4$H$_\mathrm{x}$ ions through reactions with neutral hydrocarbons. Among these, reactions involving C$_2$H$_\mathrm{x}$ ions (x= 2-6) with unsaturated hydrocarbons such as C$_2$H$_2$ were found to be more efficient that those with saturated hydrocarbons such as methane and ethane. The larger hydrocarbon ions are then lost through dissociative recombination around 320\,km.

\citet{Kim1994} note that carbon-adding reactions may continue until the carbon-growing hydrocarbon ion recombines, leading to the production of larger neutral hydrocarbons, some of which could condense to form hazes due to their low saturation vapor pressure. However, detailed calculations of the heavier hydrocarbon production rate require a large chemical network.

\citet{Perry1999} expanded on this approach and added the effect of precipitating auroral electrons on the hydrocarbon ion chemistry. They calculated the density profile for 21 ions and neutral species, and the electronic transport was performed using a multistream electron-transport code. Mono-energetic electron beams were assumed with energies ranging from 20-100\,keV with their energy flux calibrated to produce 60\,kR of H$_2$ Lyman-band emission, in agreement with HST observations of \citet{Kim1997}. The mixing ratio of stable hydrocarbons such as CH$_4$, C$_2$H$_4$, and C$_2$H$_6$ were held constant below the 4-5\,$\mu$bar level, based on Voyager stellar occultation measurements as well as photochemical calculations from \citet{Gladstone1996}. In order to account for the location of the methane homopause at higher altitude in the auroral regions, they tested three model atmospheres, with homopause altitudes ranging from 375 to 505\,km. 

Ion production peaks near the electron energy deposition peak, located around 300\,km, and depends significantly on both the assumed homopause altitude and precipitating electron energy. \cite{Perry1999} found that the major ions produced are transformed by ion-molecule reactions, with production through direct electron impact playing a smaller role. About 50\,km below the H$_3^{+}$ density peak, i.e., 400-475\,km depending on the model parameters, lies the CH$_5^{+}$ density peak, produced from the reaction between CH$_4^{+}$ and H$_2$. In between both density peaks lies the CH$_3^{+}$ density peak, though predicted to be two orders of magnitude smaller than that of CH$_5^{+}$. Overall, the ion density profiles were similar to those calculated by \citet{Kim1994}, but with abundance peaks one order of magnitude greater in the auroral region.


Combining the extended neutral chemical network from \citet{Gladstone1996} with the ion-neutral chemical reactions from \citet{Perry1999}, \citet{Wong2000} developed a model to quantitatively assess the formation of benzene (C$_6$H$_6$) and polycyclic aromatic hydrocarbons (PAH). They included nearly a hundred reactions to simulate the formation of PAHs, up to pyrene (C$_{16}$H$_{10}$), using a chemical scheme proposed in flame chemistry experiments, where the abundance of PAHs and efficiency of soot formation can be explained under the H-abstraction-C$_2$H$_2$ mechanism \citep[e.g.][]{Wang1994}. They predicted the peak abundance of several PAHs to be located in the 0.1-0.01\,mbar, and demonstrated that, by turning off the ion-neutral reactions, the predicted benzene concentration was decreased by a factor of 550.

Chemical calculations demonstrate that the production of benzene and aromatic ring species are increased in Jupiter's high-latitude auroral environment. The microphysical processes from which these aromatic species subsequently form aerosols was studied by \citet{Friedson2002}. Using the model output from \citet{Wong2000}, \citet{Friedson2002} implemented an aerosol formation microphysical model that controls their nucleation, growth, and sedimentation, based on numerical tools developed by \citet{Toon1988}. Using the PAH naming convention from \citet{Wang1994}, where A$_i$ represents an aromatic molecule of $i$ fused rings (e.g., A$_1$, A$_2$, and A$_4$ are respectively benzene, naphthalene, and pyrene), they estimated that homogeneous nucleation of the larger PAHs considered (A$_4$) become important near 0.2\,mbar through Brownian coagulation due to their high saturation ratios. The authors however noted the lack of experimental data to compute the homogeneous nucleation rate of A$_3$ and A$_4$ PAHs. These are important because they then serve as condensation nuclei through the more efficient heterogeneous nucleation for the lighter PAHs as they sediment down into the lower stratosphere. 

The vertical distribution of the produced aerosols as a function of their size then results from an interplay between the sedimentation and the coagulation removal timescales. Above the 1-mbar pressure level, the sedimentation time of small particles is short compared to their coagulation one. Below the 1-mbar pressure level, the coagulation timescale becomes shorter than the sedimentation time and particles coagulate into larger ones. \citet{Friedson2002} noted the importance of the fractal nature of the particles, as the fall time of fractal aggregates is longer than for spherical particles. This leads to the production of larger particles at a given pressure level in the case of fractal aggregates. At 20\,mbar and using two extreme fractal natures of the produced aerosols (spherical and of fractal dimension of 2.1), they predicted the mean particle radius to be between 0.1-0.7\,$\mu$m.

An indirect proof of the aerosol growth with depth in Jupiter's auroral regions was suggested by \citet{Cavalie2023b} to explain the HCN depletion observed at polar latitudes and pressure levels $>$ 0.1\,mbar (Figure \ref{fig:HCN_vs_aerosol}). They hypothesized that HCN bonded onto the mid-sized aerosols modeled by \citet{Friedson2002}. Laboratory work led by \citet{Perrin2021} under Titan conditions in terms of composition (95\% N$_2$ - 5\% CH$_4$ mixture) demonstrated that ionizing radiation first leads to the formation of small aerosols. These eventually bond with HCN as soon as they reach the size modeled by \citet{Friedson2002} in the 0.1-1 mbar region. 

The observed distribution of polar aerosols, combined with the observed depletion of HCN, suggests that they are both affected by the polar atmospheric dynamics. We now briefly discuss observational constraints of the upper atmosphere dynamics.

\section{Magnetosphere-Ionosphere Coupling: consequence for stratospheric dynamics}
\label{s:coupling}

Jupiter's volcanically-active moon Io is the major source of plasma within the Jovian magnetosphere. Over a ton per second of sulfur dioxide escapes from the moon as a neutral cloud around the moon. High-energy electrons trapped in Jupiter's magnetosphere dissociate and ionize the neutral material, producing ions which get picked up by Jupiter's strong rotating magnetic field, forming the Io plasma torus \citep{Thomas2004, Bagenal2020}.

\begin{figure}[!h]
 \begin{center}
    \includegraphics[width=0.99\textwidth]{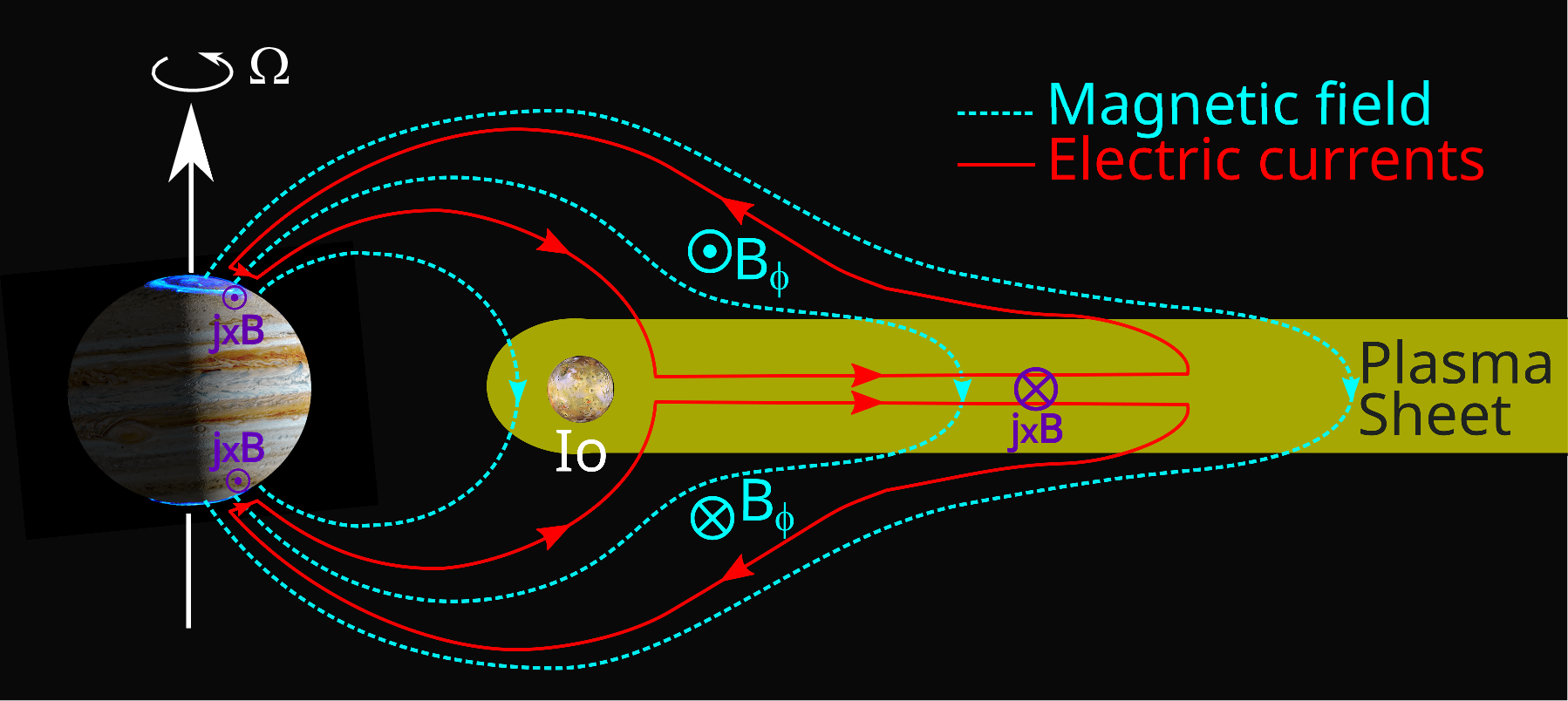}
  \end{center}
  \caption{Magnetosphere-ionosphere coupling at Jupiter following the corotation-enforcement theory. Adapted from \cite{Cowley2001} and \cite{Bonfond2020}. Field-line currents develop in the middle-magnetosphere (20-60\,R$_J$) in response of the magnetodisk plasma lagging behind corotation, resulting in an exchange of momentum with the Jovian ionosphere.}
  \label{fig:Bonfond2020}
\end{figure}

The plasma originating from Io expands to the outer magnetosphere and its azimuthal speed is expected to decrease with radial distance, based on the conservation of momentum. As it happens, there is a differential velocity between the flux tube, magnetically connected to the magnetospheric plasma, and the upper atmosphere of Jupiter, which exchange angular momentum through the \textbf{j} $\times$ \textbf{B} force (Fig. \ref{fig:Bonfond2020}). This results in spinning up the magnetospheric plasma closer to co-rotation with the planetary interior, and slowing down the Jovian ionosphere below the planetary rotation rate \citep{Hill1979, Hill2001, Cowley2001}.

Magnetosphere-ionosphere-thermosphere (MIT) models have previously been developed to investigate how such coupling affects the Jovian upper atmospheric thermal structure and dynamics \citep[e.g.][]{Achilleos1998, Bougher2005, SmithAylward2009, Tao2009, Yates2012, Ray2015}. Since the focus of this paper is on the stratospheric chemical distributions, more extensive discussion on this topic is deferred to a previous review paper \citep[e.g.][]{Bougher2008}. However, we highlight the observational proofs of this momentum exchange.

In what follows, we use the term ion and neutral winds for winds measured on ion species (H$_3^+$) and neutral species (H$_2$, HCN, CO, etc.). The first evidence of the ionosphere circulation was observed in H$_3^+$ using CSHELL, a near-IR echelle spectrograph mounted at the IRTF \citep{Rego1999}. They measured the Doppler shift of the 3.953\,$\mu$m H$_3^+$ emission line in the auroral region induced by supersonic winds in Jupiter upper stratosphere. They found counter-rotating winds, i.e., flowing opposite to the planetary rotation, in both hemispheres. From an observer's point of view, this corresponds to a clockwise wind around the polar region. Ion winds were found near Jupiter's main auroral oval, with velocities of 2.7$\pm$0.3 and 2.9$\pm$0.3\,km/s in the northern and southern hemispheres, respectively. 

This work was extended by \citet{Stallard2001, Stallard2003}, using additional CSHELL observations of the same spectral line. They found velocities of the electrojet around the main oval in the 0.5-1.5\,km/s range, as measured from positioning the slit over several cross-section across the auroral region. \citet{Stallard2003} transformed these velocities into a reference frame where the rotation of the magnetic pole was set to zero, and identified a region located within the polar aurora that was close to stationary, which was termed the fixed dark polar region. Near-stagnation in the ionosphere in the rest frame of the magnetic pole suggests a coupling to the solar wind. However, it is unknown whether the coupling is through a Dungey-like single-cell open field and return flow \citep{Cowley2003b} or Kelvin-Helmholtz instabilities in viscous flow interactions on the dawn flank \citep{Delamere2010}.


\citet{Chaufray2011} used observations made with the Fourier Transform Spectrometer instrument at the Canada-France-Hawaii-Telescope to measure the thermospheric wind velocities on two different species; H$_2$ and H$_3^+$. They used the H$_2$ S$_1$(1) quadrupole line at 2.122\,$\mu$m (4712.9\,cm$^{-1}$), and the H$_3^+$ 2\,$\nu_2$ R(7,7) line at 2.113\,$\mu$m (4732.5\,cm$^{-1}$) to derive the velocities from both ion and neutral species in the northern auroral region. The measured wind field in the auroral region was $\sim$3.1\,$\pm$0.4\,km/s for the ion wind from H$_3^+$ and they derived an upper limit for the neutral wind velocity of 1\,km/s from H$_2$.

Additional observations of the 3.953\,$\mu$m H$_3^+$ line were subsequently recorded using the Cryogenic Infrared Echelle Spectrograph (CRIRES) instrument at the European Southern Observatory Very Large Telescope (VLT). \citet{Johnson2017} derived the ion wind field in the northern auroral region at higher spatial resolution. Similar to \citet{Rego1999} and \citet{Stallard2001} they found a counter-rotating wind located near the main oval flowing at around 1.5\,km/s on the dusk side of the main oval, and 2.2\,km/s in a region poleward of the main oval on the dawn side. Errors on the measured winds were estimated to be $\pm$\,0.3\,km/s. \citet{Johnson2017} measured super-rotating winds flowing equatorward of the main oval with velocities of $\sim$\,0.7\,km/s, also previously identified by \citet{Rego1999} and \citet{Stallard2001}.

\begin{figure}[!h]
 \begin{center}
    \includegraphics[width=0.9\textwidth]{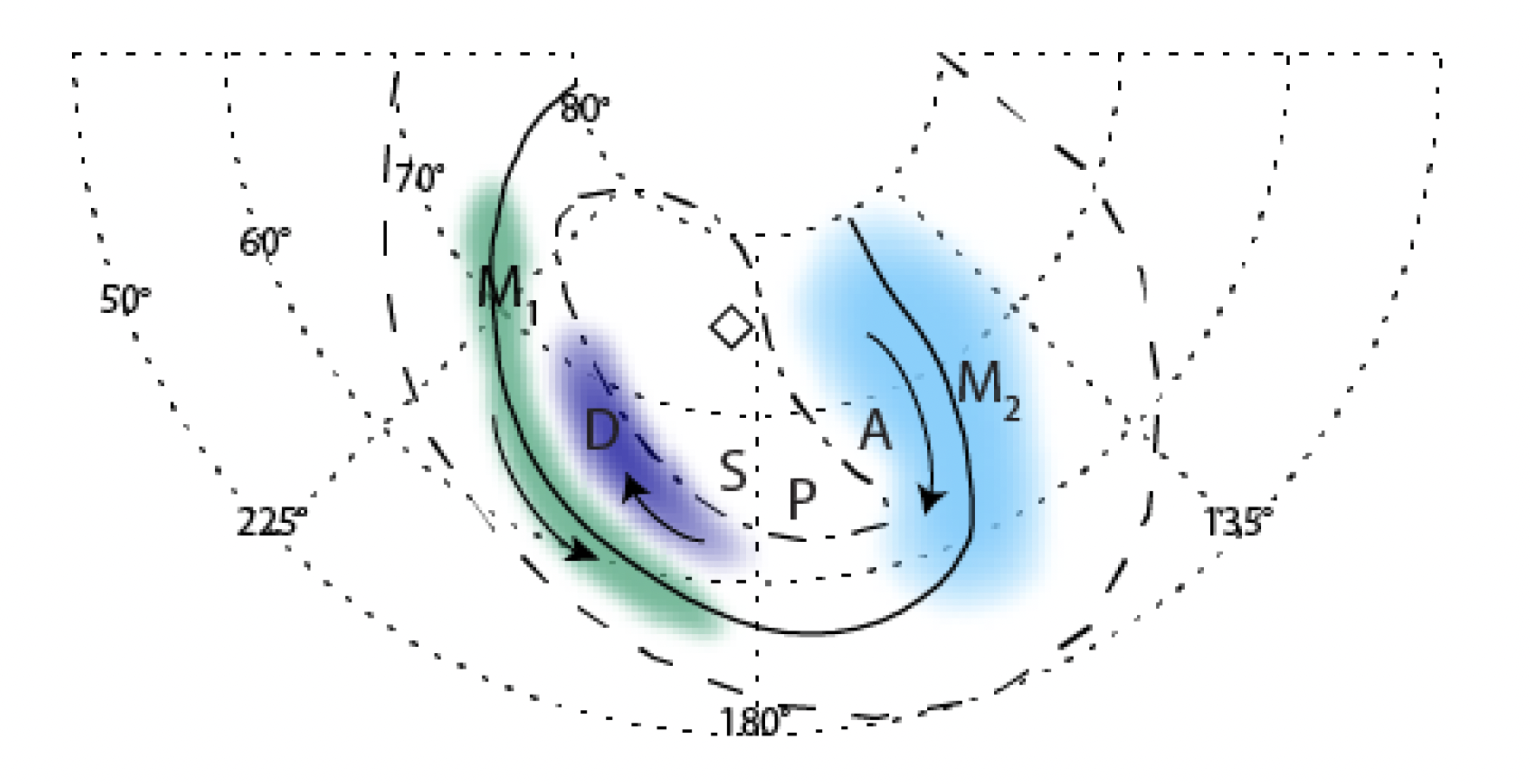}
  \end{center}
  \caption{Ionospheric flow measured around the Jovian north pole and shown as a north polar projection. The M$_2$ and D ionospheric winds are counter-rotating, while M$_1$ is super-rotating. The solid black line represents the position of the peak H$_3^+$ intensity. The diamond represents the center position of the aurora, as defined by \citet{Grodent2003}, the dashed line represents the Io footprint location from model of \citet{Grodent2008}. The dot-dash line marks the location of the fixed dark polar region, as defined by \citet{Stallard2003}. Figure from \citet{Johnson2018PhD}.}
  \label{fig:JohnsonThesis}
\end{figure}

\citet{Wang2023} recently observed the quadrupole H$_2$ line, as well as the overtone H$_3^{+}$ lines using the Keck II/NIRSpec instrument, allowing the simultaneous derivation of the ion and neutral winds for the first time. The measured ion wind field is consistent with the past H$_3^{+}$ monitoring campaigns of \citet{Stallard2003} and \citet{Johnson2017}, with globally counter-rotating jets on the dusk sector and in the dawn polar region. By subtracting the neutral winds measured on H$_2$ from the ionospheric winds measured in H$_3^{+}$, \citet{Wang2023} derived the effective ion winds for the first time, providing critical constraints on how the ion winds progressively drag the neutral atmosphere with increasing pressure.

\begin{figure}[!h]
 \begin{center}
    \includegraphics[width=0.97\textwidth]{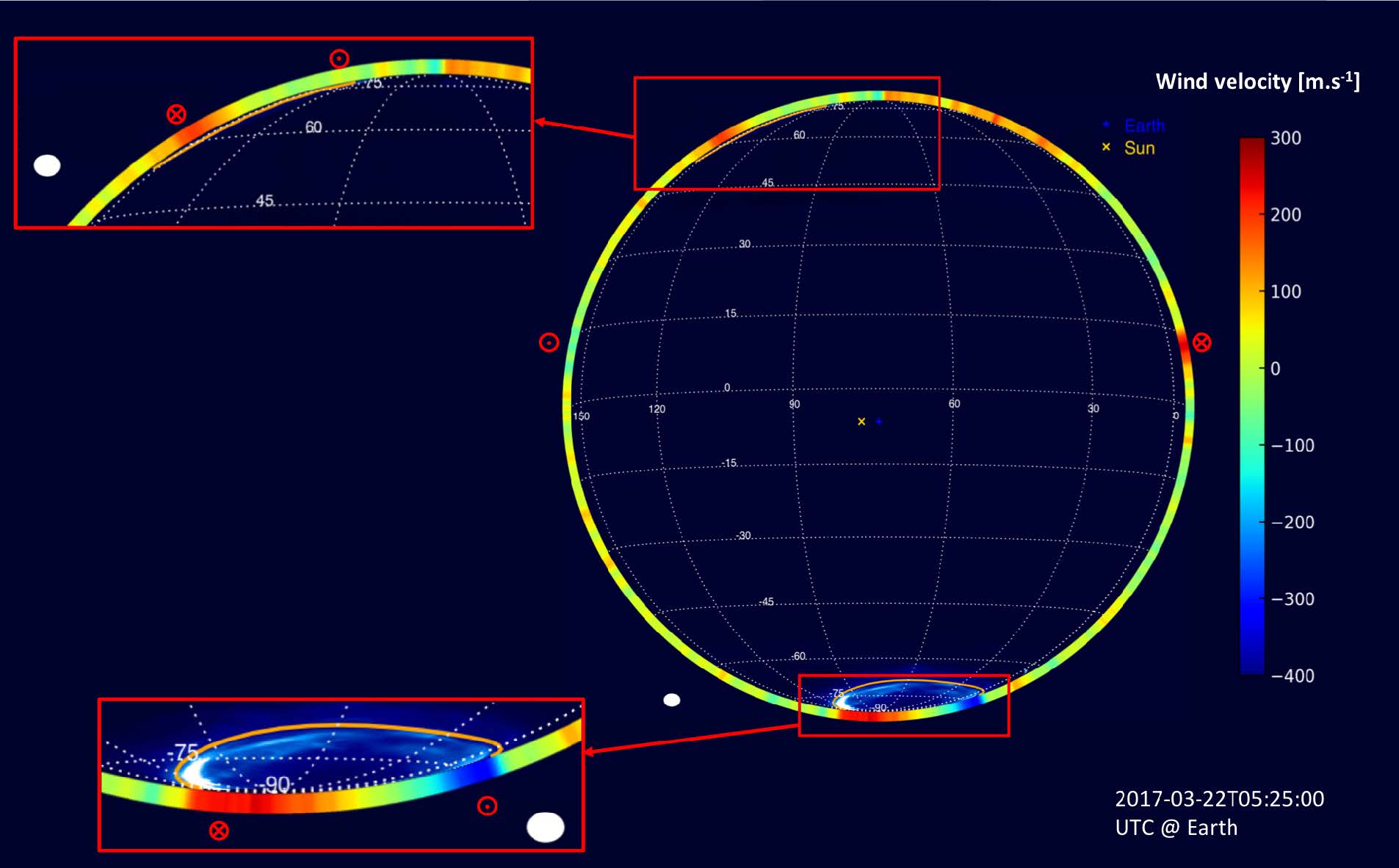}
  \end{center}
  \caption{Jupiter’s UV aurora and stratospheric winds. This composite image shows the line-of-sight (LOS) wind velocities (in m\,s$^{-1}$) derived from the ALMA observations and the statistical emission of the aurorae \citep{Clarke2009} in the configuration of the ALMA observations. The northern and southern aurora regions are best seen in the dedicated zoomed-in quadrants. The M = 30 footprints of the magnetic field model from \cite{Connerney2018} are good markers of the positions of the main ovals as seen by Juno-UVS \citep{Gladstone2017_SSR} and are plotted in orange. The white ellipses indicate the spatial resolution of the ALMA observations. The directions of the strongest winds in the equatorial and auroral regions are indicated with the red $\odot$ and $\otimes$ symbols. Image reproduced from \cite{Cavalie2021}, copyright by the author(s).}
  \label{fig:ALMA_cavalie2}
\end{figure}

One last piece of the puzzle was brought by \citet{Cavalie2021}, who targeted several post SL9-species (HCN and CO) using ALMA. Figure \ref{fig:ALMA_cavalie2} shows the measured wind-induced Doppler shift measured on the HCN (4-3) line at 354.505 GHz. This dataset provides neutral wind information for the first time in Jupiter's stratosphere at a pressure level of 0.1\,mbar at high latitudes, and around 3\,mbar at lower latitudes.

\citet{Cavalie2021} measured counter-rotating neutral winds flowing in the southern polar region, with velocities of 350\,$\pm$\,20\,m/s around the dusk side of the main oval (eastern limb) and 200\,$\pm$\,20\,m/s around the dawn side of the main oval (western limb). They also detected hints of the northern hemisphere wind pattern counterpart, but because the observation geometry of the northern polar region was not favorable at that time, additional measurements of the northern polar region are needed to fully characterize the neutral wind regime in that hemisphere.




\begin{figure}[!h]
 \begin{center}
    \includegraphics[width=0.9\textwidth]{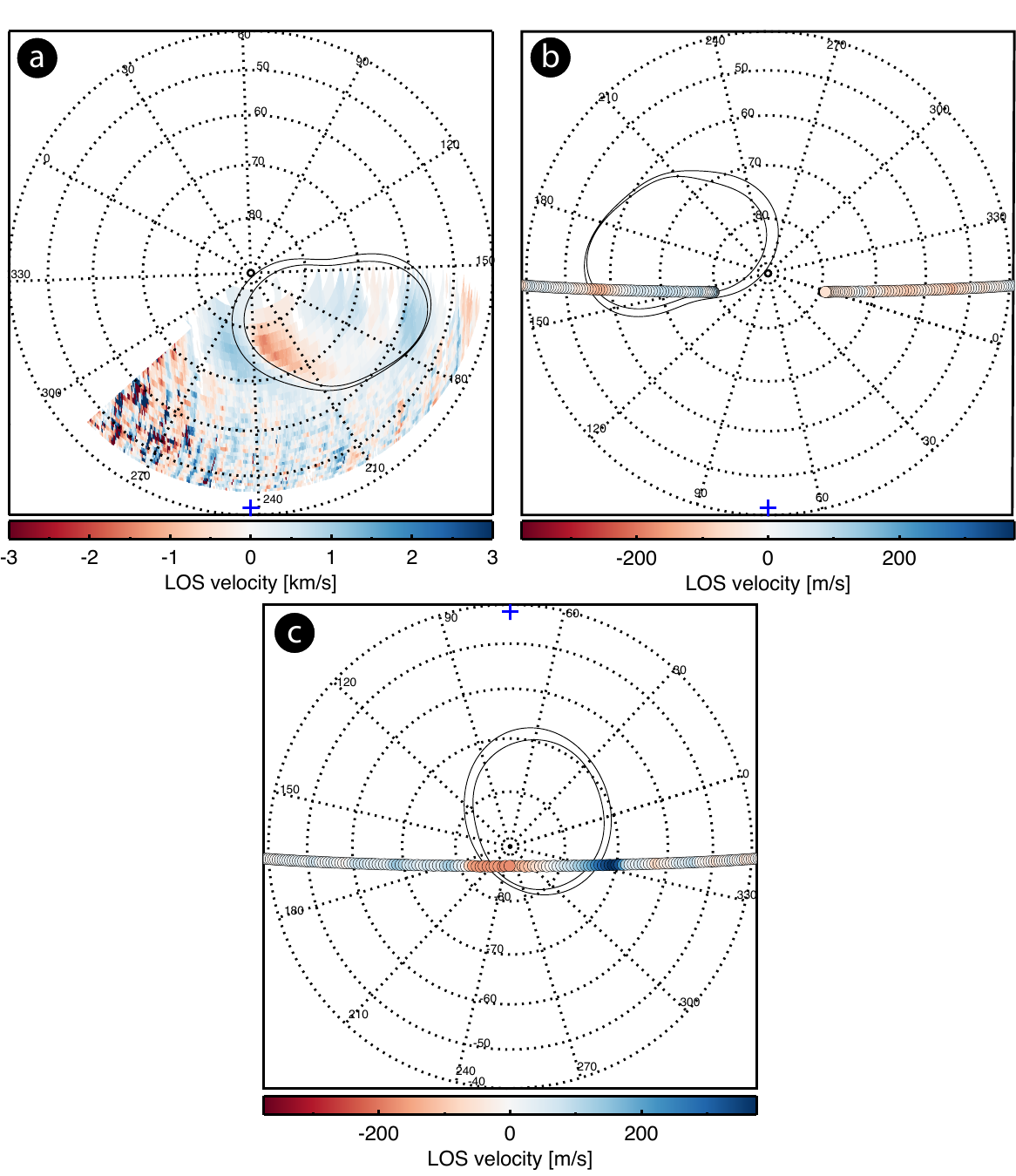}
  \end{center}
  \caption{Panel a: northern electrojet wind measurements on H$_3^{+}$ recorded at CML of 242$^{\circ}$W \citep{Johnson2017}. Panel b and c: neutral stratospheric wind measurements on HCN from ALMA at CML $\sim$73$^{\circ}$W in the northern and southern hemisphere, respectively \citep{Cavalie2021}. The observation central meridian longitudes are indicated as blue crosses. All winds are expressed in a reference frame that is rotating with the planet.}
  \label{fig:winds_ALMA_h3p}
\end{figure}

Figure \ref{fig:winds_ALMA_h3p} compares the ionospheric wind derived from H$_3^{+}$ observations performed with VLT/CRIRES on 31 December 2012 (upper left panel) with the stratospheric neutral wind derived from HCN observations performed by ALMA on 22 March 2017 (upper right and bottom panels). The northern ionospheric and stratospheric winds were recorded at Central Meridian Longitudes (CML) almost half a Jupiter rotation apart (CML of 242$^{\circ}$W and 73$^{\circ}$W, respectively), though in different years. These observations suggest that the equatormost part of the electrojet (Fig. \ref{fig:winds_ALMA_h3p}a at longitudes 150$^{\circ}$W-180$^{\circ}$W and latitudes 55$^{\circ}$N-60$^{\circ}$N) might be stable over time and potentially extends down to the neutral atmosphere at least to sub-mbar levels (Fig. \ref{fig:winds_ALMA_h3p}b at longitudes 150$^{\circ}$-165$^{\circ}$ and latitude 55$^{\circ}$N-60$^{\circ}$N), although the detailed mechanism needs to be worked out. This conclusion can also be derived from observations of \citet{Chaufray2011} and \citet{Wang2023} indicating the co-location of ion and neutral winds, though derived from slightly different altitudes. The southern polar region stratospheric winds are more obvious, though additional VLT/CRIRES observations of the southern hemisphere would be needed to reinforce this conclusion.



The neutral wind measurements derived from HCN and H$_2$ across the polar stratosphere do not indicate super-corotating flows, while the ion wind measurements consistently do \citep{Rego1999, Stallard2001, Johnson2017, Wang2023}. Morphology of the sub-rotating neutral winds from \citet{Wang2023} match fairly well the ion winds one measured on the same dataset, without the super-rotating ion winds. Models of this interaction suggest the neutral atmosphere is slow to react to changing ion winds with ion drag. Notably, \citet{Tao2009} showed that acceleration of the neutral atmosphere caused by ion drag decreases with increasing altitude. Using calculations from \citet{Tao2009}, \citet{Wang2023} estimated that this process takes about 10 Jovian days to accelerate the neutral atmosphere to $\sim$1 km/s at an altitude of 1000\,km, while it takes about 70 Jovian days to do so at 400\,km altitude. As a result, while H$_3^+$ takes on any asymmetric flows in the magnetosphere, e.g. from super-rotation as measured on the dayside magnetosphere \citep{Krupp2001} and modeled \citep{Chane2013}, the neutral atmosphere probably does not.

\begin{figure}[!h]
 \begin{center}
    \includegraphics[width=0.9\textwidth]{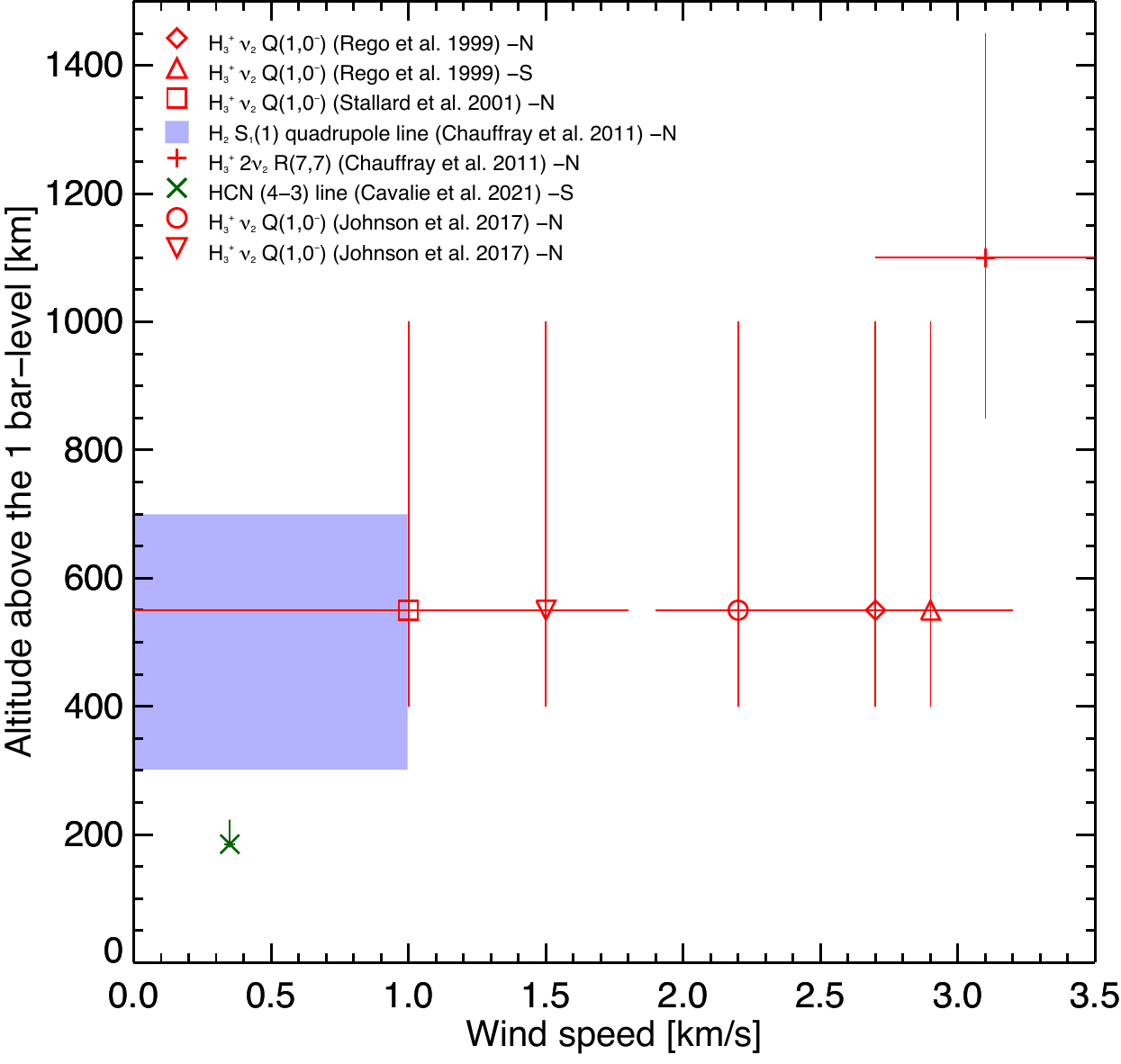}
  \end{center}
  \caption{Summary of the ionospheric and stratospheric wind measurements in Jupiter's stratosphere. The label describes the spectral information of the line, as well as the hemisphere (N/S) targeted. Information about the altitude sensitivity of these measurements are derived from \citet{Melin2005} and \citet{Grodent2001}. Wind measurements from \citet{Chaufray2011} and derived from the H$_2$ S$_1$(1) quadrupole line provided an upper limit of 1\,km/s, over the altitude range of 300–700\,km (see text for details).}
  \label{fig:Summary_Stratospheric_Wind}
\end{figure}

Figure \ref{fig:Summary_Stratospheric_Wind} summarizes the aforementioned ionospheric and stratospheric neutral wind measurements. The altitude region at which the H$_3^+$ observations are sensitive to are taken from \citet{Melin2005}, which uses the atmospheric model of \citet{Grodent2001}. The altitude range of these H$_3^+$ observations are estimated from the FWHM of the line emission intensity. For instance, the altitude sensitivity of the fundamental $\nu_2$ Q(1,0$^{-}$) line is 550\,$^{+450}_{-150}$\,km ($\sim$ 4$\times$10$^{-4}$-4$\times$10$^{-6}$\,mbar). We assumed the altitude at which the 2$\nu_2$ R(7,7) line senses to be 1100\,$^{+350}_{-250}$\,km ($\sim$ 1$\times$10$^{-5}$-3$\times$10$^{-7}$\,mbar), based on the altitude sensitivity range of the 2$\nu_2$ lines calculated by \citet{Melin2005}. The pressure range at which the HCN (4-3) line forms is 0.08\,$^{+0.02}_{-0.06}$\,mbar, assuming the peak and the FWHM of the line forming contribution function from \citet{Cavalie2021}. This corresponds to an altitude range of 185\,$^{+38}_{-5}$\,km, using the \citet{Grodent2001} model. The altitude sensitivity range of the H$_2$ S$_1$(1) quadrupole line was calculated to be 300-700\,km, based on \citet{Chaufray2011} and \cite{Bougher2005}. Note that additional models of \citet{Tao2011} predict sensibly similar peak UV (H$_2$) and IR (H$_3^+$) emission altitudes.

\section{Summary and outstanding questions}
\label{ss:summary}

Observations of the Jovian upper atmosphere in the UV, IR and mm/sub-mm all provide evidence that the chemical distributions and thermal structure are broadly influenced by auroral particle precipitation. 

Mid-IR observations of the chemical distributions suggest that several light hydrocarbons (C$_2$H$_2$, C$_2$H$_4$, C$_2$H$_6$) are locally affected near the statistical location of the main auroral region. The derived C$_2$H$_2$ and C$_2$H$_4$ abundances are enhanced at the mbar and sub-mbar levels consistently in all mid-IR datasets. The situation for C$_2$H$_6$ is interesting as earlier space-based observations suggested it was depleted within the auroral oval at 5\,mbar, but enhanced at the same latitude away from the oval, implying some chemical conversion, as the total C$_2$H$_\mathrm{x}$ content seems to be zonally conserved at the mbar level \citep{Sinclair2017}. Recent datasets recorded at high spatial resolution confirm that C$_2$H$_6$ is enhanced at high latitudes \citep{Sinclair2023, RodriguezOvalle2024}, and likely affected by modulation in the magnetospheric activity, possibly linked with changes in the solar wind conditions \citep{Sinclair2023}.

Ultraviolet observations with Juno-UVS have provided another important piece of the puzzle. Solar-reflected UV light captured by UVS around 175-190\,nm carry the absorption signature of C$_2$H$_2$ and C$_6$H$_6$, probing their total column abundance down to 5-50\,mbar \citep{Giles2023}. Juno-UVS observations show that C$_2$H$_2$ is enhanced up to a factor of 3.5 around the polar auroral region, compared to the non-auroral atmosphere, and possibly suggest that C$_6$H$_6$ is enhanced around the same region. Since these observations probe deeper than the mid-IR observations and reflect the total column abundance, the enhancements are more homogeneous across the polar auroral regions.


ALMA is an outstanding tool for millimeter observations of Jupiter and can provide chemical distributions \citep{Cavalie2023b}, as well as auroral \citep{Cavalie2021} and equatorial winds \citep{Benmahi2021} in a single observation run. One of the most striking features from these datasets is the almost 2 orders of magnitude drop in the HCN column density above both poles, initially observed by Cassini-CIRS over the south pole \citep{Lellouch2006}. ALMA observations provide a detailed mapping of this depletion, and demonstrate that it is located at pressures where large auroral-produced aerosols \citep{Friedson2002} are expected. \citet{Cavalie2023b} even suggest that HCN could have an auroral production term at higher altitude.

Neutral photochemical models of the stratosphere have attempted to reproduce the meridional trends in the C$_2$H$_2$ and C$_2$H$_6$ observed by Cassini-CIRS and retrieved by \citet{Nixon2007, Nixon2010}, which initially excluded the longitudinal range covering the auroral oval. The anti-correlated trend in both hydrocarbon abundances (C$_2$H$_2$ decreasing towards the pole, while C$_2$H$_6$ increases towards the pole) still remains unexplained. It was initially thought that, because both hydrocarbons have a different lifetime, there may be a circulation pattern that could produce such trend. Using a coupled photochemical-transport 2D-model, \citet{Hue2018} showed that both trends can not be reproduced by a combination of advective and diffusive transport, and that another chemical process must be preferentially affecting one of these hydrocarbons.

Ion-neutral chemical models provide insights on how the charged particle precipitation affect the neutral species. \citet{Kim1994} predicted that CH$_5^{+}$, C$_2$H$_3^{+}$, CH$_3^{+}$ and C$_2$H$_5^{+}$ are the dominant hydrocarbon ions at pressures of 7$\times$10$^{-4}$\,mbar and above in the upper atmosphere. They showed that reactions involving C$_2$H$_\mathrm{x}$ ions with unsaturated hydrocarbons such as C$_2$H$_2$ are more efficient than with saturated hydrocarbons (e.g., CH$_4$ and C$_2$H$_6$). These could possibly explain the trend seen by \citet{Sinclair2017} where C$_2$H$_2$ is locally enhanced around the polar auroral region, and converted into C$_2$H$_6$ away from that region, although \citet{Kim1994} did not study how the auroral electron precipitations would affect the neutral hydrocarbon abundances. They noted that the combination of hydrocarbon ions could lead to the formation of larger neutral hydrocarbons. 

The link between the formation of large hydrocarbon molecules with the observation of the polar hazes on Jupiter has been established for years \citep{Hord1979, Pryor1991}. Work by \citet{Zhang2013b} using Cassini/ISS limb-darkening observations at multiple phase angles showed the existence of two types of aerosols; one consisting of compact sub-micron particles, and the other consisting of fractal aggregates composed of small monomers. The high-latitude regions were best reproduced assuming fractal aggregated particles consisting in thousands of $\sim$10\,nm monomers, with column densities $\sim$2 orders of magnitude greater than around the low-latitude regions. \citet{Wong2000, Wong2003} and \citet{Friedson2002} estimated the amount of large hydrocarbons produced from ion-neutral reactions, and developed a microphysical model of the auroral haze formation. Hydrocarbons are produced from methane chemistry and ion-neutral chemistry around the nbar level, they are predicted to grow into larger hydrocarbon fused rings around 10\,nbar. They may subsequently nucleate through homogeneous nucleation at the submbar level, and grow through heterogeneous nucleation while sedimenting. At the mbar level, they coagulate and condensate into sub-microns aerosols.

One important piece from two decades of H$_3^+$ observations as well as ALMA observation concerns the dynamical coupling of the upper atmosphere with the magnetosphere, potentially affecting the chemical spatial distribution. H$_3^+$ observations provided evidence of an ionospheric auroral electrojet \citep{Rego1999, Stallard2001, Johnson2017, Wang2023} flowing near the main oval opposite to the planetary rotation, and may extend to the neutral atmosphere down to the sub-mbar level, i.e., several hundred km below the electrojets. 

These electrojets corotate with Jupiter at a rate of 10 hours, while the tropospheric weather layer does not rotate exactly at that rate. There is a progressive dynamical decoupling operating between (i) the ionospheric vortex flowing along the main oval, and (ii) the background neutral atmosphere underneath. Depending on the pressure level probed, different trends are expected for the chemical distributions. At the nbar to mbar levels, aurorally-produced species may be confined within the main oval as a result of short chemical lifetimes possibly combined with the dynamical confinement caused by the electrojet. This seems to be what the hydrocarbon distributions sensed by mid-IR observations are suggesting \citep[e.g.][]{Sinclair2018, Sinclair2023}. At the 10-100 mbar level, the situation is opposite and the chemical distributions are longitudinally homogeneous, as seen, e.g., in the polar haze distributions \citep{Zhang2013b, Rogers2022, Hueso2023}, also shown on Fig. \ref{fig:polar_methane_band}. In between these two pressure levels, chemical distributions are expected to display a mixture of both morphologies. This may be the case for the Juno-UVS reflected sunlight observations in the southern hemisphere (Fig. \ref{fig:Giles2023_vs_Cavalie2023}), which shows a C$_2$H$_2$ enhancement within the main oval, superposed with a longitudinally-homogeneous background enhancement that mimics the polar haze distributions \citep{Giles2021b}. 



\begin{figure}[!h]
 \begin{center}
    \includegraphics[width=0.98\textwidth]{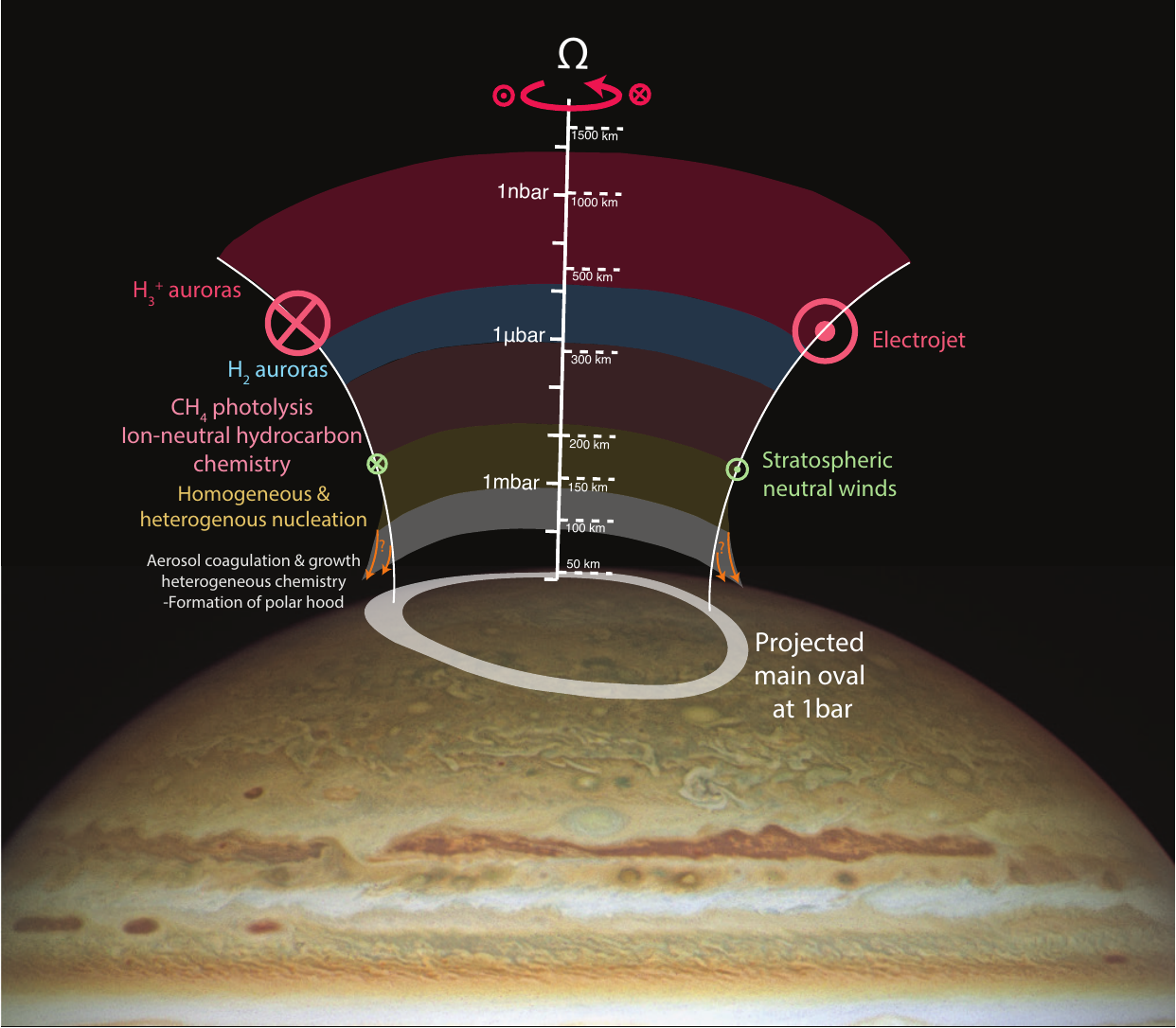}
  \end{center}
  \caption{Summary schematics of selected chemical distributions across Jupiter's polar region. Particle precipitation leads to the formation of H$_3^+$ in the ionosphere, and to UV-auroras through the H and H$_2$ emissions at the $\mu$bar level. Around the $\mu$bar level and below, solar-UV photolysis as well as charged particle precipitations leads to the formation of hydrocarbons. As hydrocarbons diffuse downward, they combine into heavier hydrocarbons to form aerosols. HCN produced in the upper atmosphere, possibly from N$_2$ destruction, get incorporated into aerosols before they reach $\sim$2\,mbar or below. Background image credits: NASA, ESA, A. Simon, M. H. Wong, and J. DePasquale.}
  \label{fig:summary}
\end{figure}

Figure \ref{fig:summary} summarizes the various processes discussed in this paper. Note that, although ALMA data have unequivocally indicated the existence of stratospheric neutral winds co-located with the position of the southern main auroral emission \citep{Cavalie2021}, it only provided hints of that circulation in the northern hemisphere due to the observational geometry. Some of the remaining outstanding questions that need to be answered include:

$\bullet$ What are the exact chemical pathways producing the enhanced hydrocarbons in the polar region, given the constraints brought by the Juno in-situ particle measurements across the polar regions?

$\bullet$ Do Jupiter's UV auroral emissions affect the high-latitude hydrocarbon distributions through photolysis?

$\bullet$ What is the origin of the double temperature peak seen in the mid-IR observations, i.e., around the 1\,mbar and 10\,$\mu$bar levels?

$\bullet$ What are the physical mechanisms causing the higher methane homopause in the polar auroral region?

$\bullet$ Where does the decoupling between the electrojet and the lower stratosphere occur? How does this decoupling affect the chemical distributions?

$\bullet$ Given the larger tilt in the northern magnetic dipole compared to the southern one, how different is the decoupling of the electrojet and the background neutral atmosphere between both hemispheres? In other words, is the southern auroral circulation more efficient at dynamically confining the aurorally-produced produced species and hazes?

$\bullet$ Does the super-rotating electrojet found equatorward of the northern main oval propagate down to the stratosphere?

$\bullet$ What is the pattern of the residual circulation in the polar region, and how does it affect the distribution of chemical species?

$\bullet$ How do the polar aerosols produce the complex pattern of wavy structures at different latitudes? 

$\bullet$ How often auroral events form specific patches of aerosols or concentrated chemicals associated to polar dark patches as seen in the UV \citep{Porco2003, Tsubota2024}? Are these patches set in motions by the same wind propagating from the ionosphere to the bottom of the stratosphere?

To answer some of these questions, continuing the monitoring in the near- and mid-IR, as well as well in the millimeter range is crucial. For instance, ALMA alone can provide the distributions of several key chemical species as well as the winds with a single dataset of moderate exposure time ($<$1\,hr on-source, thus enabling longitudinal coverage with limited longitudinal smearing) \citep{Cavalie2021}. Modeling effort should be pursued in accounting for the ion-neutral chemical network available in the literature in Jovian conditions \citep{Loison2015, Dobrijevic2016, Wong2000, Brown2024}, while including the knowledge of the charged particle precipitating energy flux measured by Juno. Some of the main ions predicted are CH$_5^{+}$, C$_2$H$_3^{+}$, CH$_3^{+}$ and C$_2$H$_5^{+}$ \citep{Kim1994}, and attempting to detect them will help to constrain ion-neutral photochemical models. Laboratory work is required to identify the spectral signatures of these ions and enable their detection with observatories such as JWST.

Finally, the JUICE mission will undoubtedly help to improve our understanding of the chemistry and dynamics of Jupiter's polar regions with its 3-year Jupiter tour including equatorial and inclined orbits. JUICE's comprehensive payload includes remote sensing instruments, spanning from the UV to the sub-mm, as well as in situ experiments \citep{fletcher2023}.

\bmhead{Acknowledgements}

V. Hue acknowledges support from the French government under the France 2030 investment plan, as part of the Initiative d’Excellence d’Aix-Marseille Université – A*MIDEX AMX-22-CPJ-04. French authors acknowledge the support of CNES to the Juno mission. A subset of the work presented in this paper was carried out at the Jet Propulsion Laboratory, California Institute of Technology, under a contract with the National Aeronautics and Space Administration (NASA). The material is based upon work supported by the NASA under Grant NNH20ZDA001N issued through the Solar System Observations Planetary Astronomy program. R. Hueso was supported by grant PID 2019-109467GB-I00 funded by MCIN/AEI/ 10.13039/501100011033/ and was also supported by Grupos Gobierno Vasco IT1742-22. The support of T. K. Greathouse and R. S. Giles was funded by the NASA's New Frontiers Program for Juno via contract NNM06AA75C with the Southwest Research Institute. R. E. Johnson is supported by NERC grant NE/W002914/1. C. A. Nixon was supported for his work on this paper by NASA GSFC SSED Strategic Science funding. We thank Bertrand Bonfond for providing Figure 11.

\section*{Compliance with Ethical Standards}

The authors declare that they have no conflicts of interest.

\bibliography{sn-bibliography}


\begin{appendices}

\section{Coverage of mid-IR and UV observing campaigns}

\label{sec:appendixA}

\begin{figure}[!h]
 \begin{center}
    \includegraphics[width=0.97\textwidth]{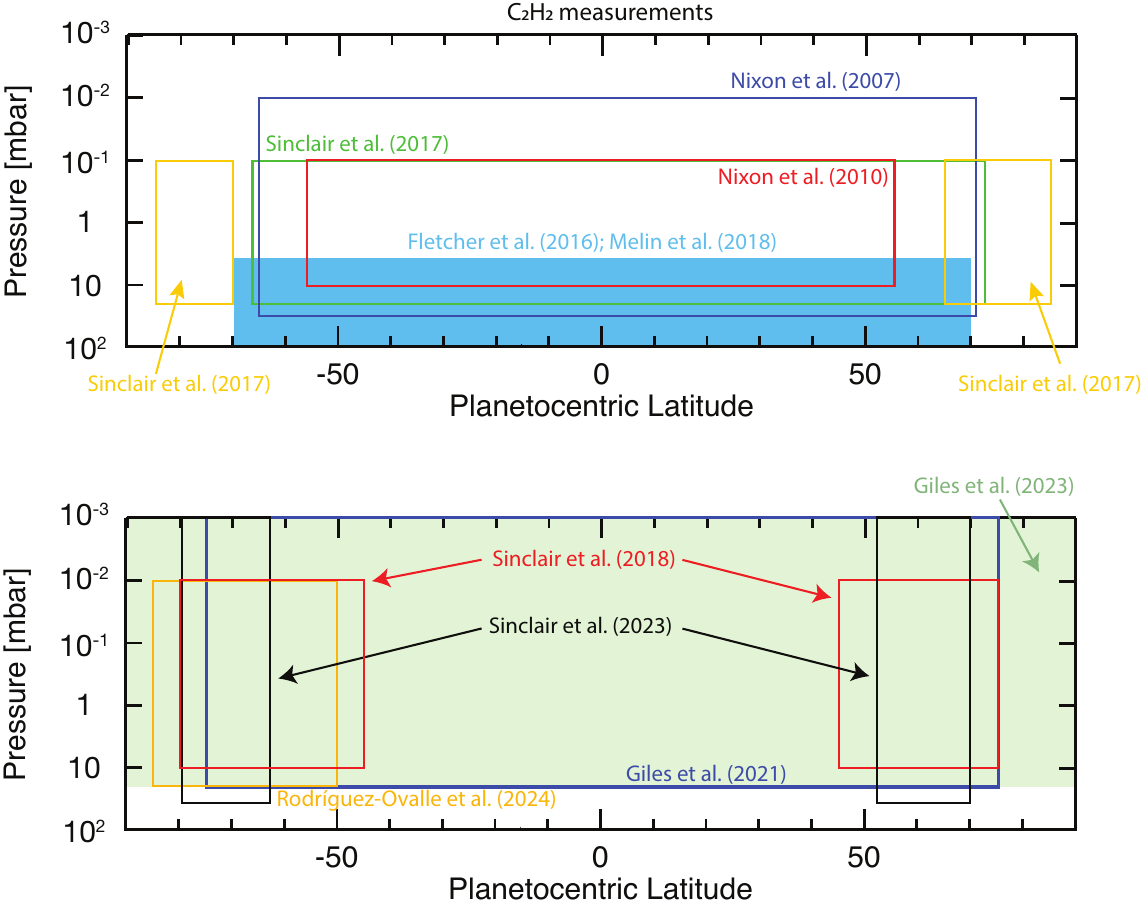}
  \end{center}
  \caption{Spatial coverage and vertical sensitivity of the mid-IR and UV observing campaigns that derived spatial distribution of acetylene (C$_2$H$_2$), from Table \ref{tab:C2H2_tab}. The coverage from these measurements were broken into two plots for clarity.}
  \label{fig:C2H2_coverage}
\end{figure}

\begin{figure}[!h]
 \begin{center}
    \includegraphics[width=0.97\textwidth]{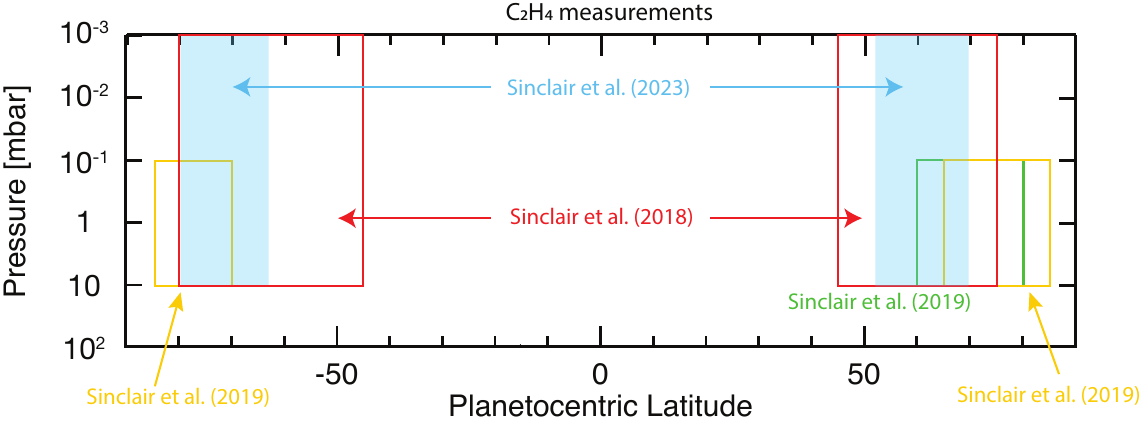}
  \end{center}
  \caption{Spatial coverage and vertical sensitivity of the mid-IR observing campaigns that derived spatial distribution of ethylene (C$_2$H$_4$), from Table \ref{tab:C2H4_tab}.}
  \label{fig:C2H4_coverage}
\end{figure}

\begin{figure}[!h]
 \begin{center}
    \includegraphics[width=0.97\textwidth]{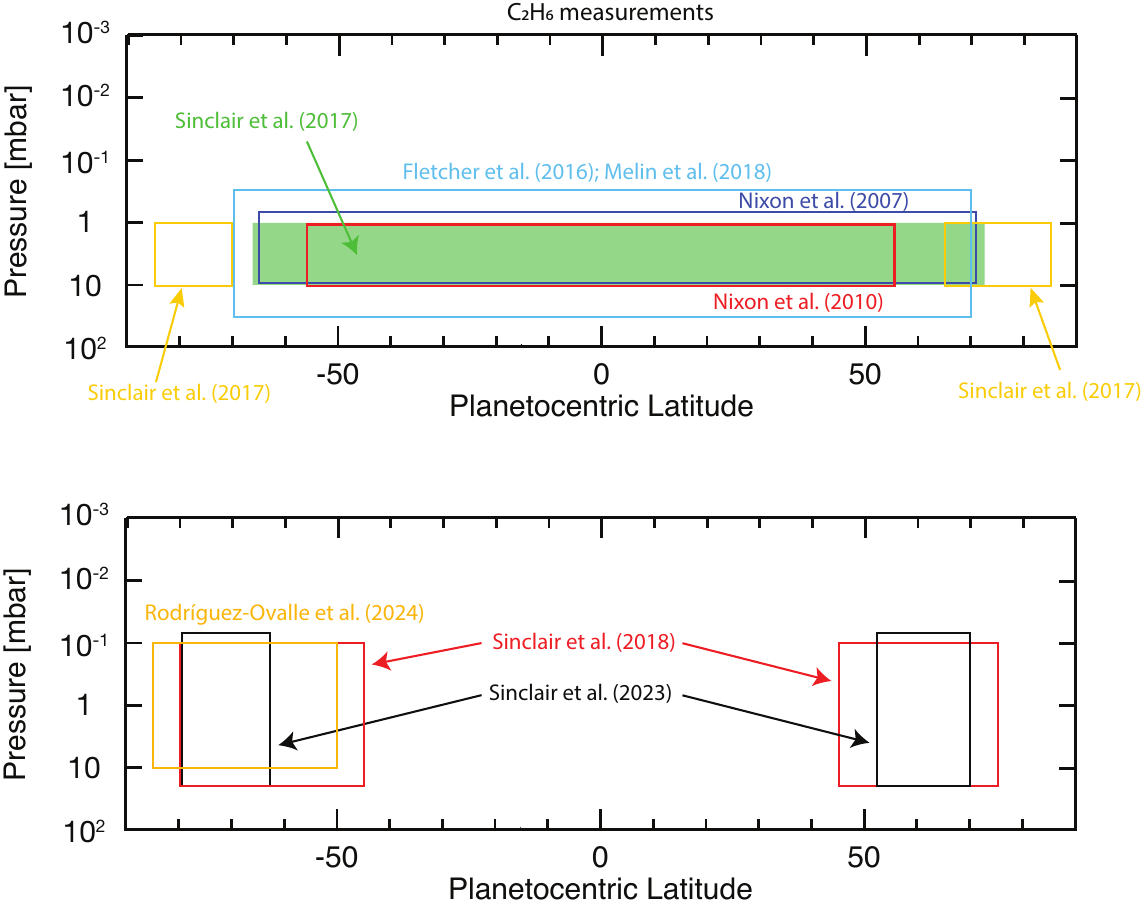}
  \end{center}
  \caption{Spatial coverage and vertical sensitivity of the mid-IR observing campaigns that derived spatial distribution of ethane (C$_2$H$_6$), from Table \ref{tab:C2H6_tab}. The coverage from these measurements were broken into two plots for clarity.}
  \label{fig:C2H6_coverage}
\end{figure}

\begin{figure}[!h]
 \begin{center}
    \includegraphics[width=0.97\textwidth]{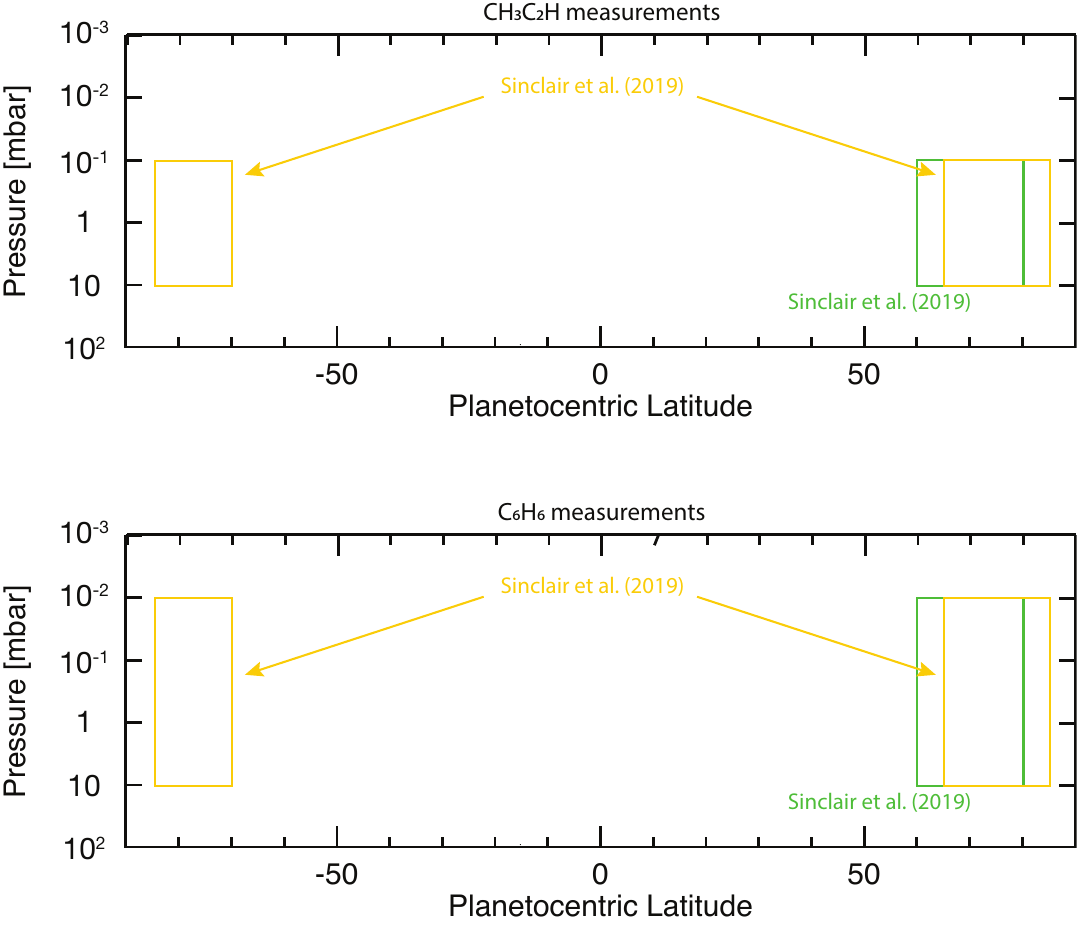}
  \end{center}
  \caption{Spatial coverage and vertical sensitivity of the mid-IR observing campaigns that derived spatial distribution of methylacetylene (CH$_3$C$_2$H, top panel) and benzene (C$_6$H$_6$, bottom panel), from table \ref{tab:CxHy_tab}. Note that JWST/MIRI provided information about the benzene distribution across the southern polar stratosphere of Jupiter \citep{RodriguezOvalle2023EGU}, but the authors concluded that these measurements were not particularly sensitive to the benzene vertical distribution.}
  \label{fig:CxHy_tab}
\end{figure}

Figures \ref{fig:C2H2_coverage}, \ref{fig:C2H4_coverage}, \ref{fig:C2H6_coverage}, and \ref{fig:CxHy_tab} represent the spatial coverage of the mid-IR and UV observations campaign from which the polar stratospheric hydrocarbon trends (tables \ref{tab:C2H2_tab}, \ref{tab:C2H4_tab}, \ref{tab:C2H6_tab}, and \ref{tab:CxHy_tab}) were derived.

Note that the vertical sensitivity of the mid-IR observations depends on the spectral setting as well as the vertical temperature profile. Observations at higher resolving power are generally sensitive to a greater pressure range. Upper-stratospheric heating, e.g. near the polar hot spot, also causes secondary sensitivity peaks in the contribution functions at higher altitudes \citep[e.g.][]{Sinclair2019}. Although Figs. \ref{fig:C2H2_coverage}, \ref{fig:C2H4_coverage}, \ref{fig:C2H6_coverage}, and \ref{fig:CxHy_tab} show the pressure range over which these observations are sensitive to, the corresponding tables only provide the trends at the pressure level at the respective contribution functions peak.

The reflected sunlight observation from Juno-UVS are sensitive to the total column abundance down to the 5-50\,mbar \citep{Melin2020, Giles2021b, Giles2023}.

\end{appendices}



\end{document}